\documentclass[5p]{elsarticle}
\usepackage[utf8]{inputenc}
\usepackage[T1]{fontenc}
\usepackage[english]{babel}
\usepackage{amsmath}
\usepackage{amssymb,amsfonts,textcomp}
\usepackage{array}
\usepackage{hhline}
\usepackage[dvipsnames]{xcolor}
\usepackage{etoolbox}
\usepackage{fixltx2e}

\colorlet{mylinkcolor}{violet}
\colorlet{mycitecolor}{YellowOrange}
\colorlet{myurlcolor}{Aquamarine}
\usepackage{epigraph}
\usepackage{graphicx}
\usepackage{enumerate}
\usepackage{newfloat}

\makeatletter
\def\blfootnote{\gdef\@thefnmark{}\@footnotetext}
\makeatother
\DeclareFloatingEnvironment[fileext=frm,placement={!htp},name=Video]{myfloat}

\journal{Progress in biophysics and molecular biology}

\date{2016-07-14}
\begin{document}

\begin{frontmatter}

\title{Modeling mammary organogenesis from biological first principles: cells and their physical constraints}

\author[msc,ihpst]{Maël Montévil}

\ead{mael.montevil@gmail.com}

\ead[url]{http://montevil.theobio.org}

\author[tufts]{Lucia Speroni}

\ead{lucia.speroni@tufts.edu}

\author[cavailles,nantes,tufts]{Carlos Sonnenschein}
\ead{carlos.sonnenschein@tufts.edu}
\author[cavailles,tufts]{Ana M. Soto\corref{mycorrespondingauthor}}
\ead{ana.soto@tufts.edu}
\cortext[mycorrespondingauthor]{Corresponding author}

\address[msc]{Laboratoire "Matière et Systèmes Complexes" (MSC),
UMR 7057 CNRS, Université Paris 7 Diderot,
75205 Paris Cedex 13, France}

\address[ihpst]{Institut d'Histoire et de Philosophie des Sciences et des Techniques (IHPST) - UMR 8590,
13, rue du Four,
75006 Paris, France}
\address[tufts]{Department of Integrative Physiology and Pathobiology, Tufts University School of Medicine, Boston, MA USA
}
\address[cavailles]{Centre Cavaillès, République des Savoirs, CNRS USR3608, Collège de France et École Normale Supérieure, Paris, France}

\address[nantes]{Institut d’Etudes Avancées de Nantes, France.}

\begin{abstract}

In multicellular organisms, relations among parts and between parts and the whole are contextual and interdependent.
These organisms and their cells are ontogenetically linked: an organism starts as a cell that divides producing
non-identical cells, which organize in tri-dimensional patterns. These association patterns and cells types change as
tissues and organs are formed. This contextuality and circularity makes it difficult to establish detailed cause and
effect relationships. Here we propose an approach to overcome these intrinsic difficulties by combining the use of two
models; 1) an experimental one that employs 3D culture technology to obtain the structures of the mammary gland,
namely, ducts and acini, and 2) a mathematical model based on biological principles.

The typical approach for mathematical modeling in biology is to apply mathematical tools and concepts developed
originally in physics or computer sciences. Instead, we propose to construct a mathematical model based on proper
biological principles. Specifically, we use principles identified as fundamental for the elaboration of a theory of
organisms, namely i) the default state of cell proliferation with variation and motility and  ii) the principle of
organization by closure of constraints. 

This model has a biological component, the cells, and a physical component, a matrix which contains collagen fibers.
Cells display agency and move and proliferate unless constrained; they exert mechanical forces that i) act on collagen
fibers and ii) on other cells. As fibers organize, they constrain the cells on their ability to move and to
proliferate. The model exhibits a circularity that can be interpreted in terms of closure of constraints. 

Implementing the mathematical model shows that constraints to the default state are sufficient to explain ductal and
acinar formation, and points to a target of future research, namely, to inhibitors of cell proliferation and motility
generated by the epithelial cells.  The success of this model suggests a step-wise approach whereby additional
constraints imposed by the tissue and the organism could be examined \textit{in silico} and rigorously tested by
\textit{in vitro} and \textit{in vivo} experiments, in accordance with the organicist perspective we embrace.
\end{abstract}

\end{frontmatter}
\blfootnote{\textbullet \ 
Published as:
Maël Montévil, Lucia Speroni, Carlos Sonnenschein, Ana M. Soto, Modeling mammary organogenesis from biological first principles: Cells and their physical constraints, \textit{Progress in Biophysics and Molecular Biology}, Available online 18 August 2016, ISSN 0079-6107, \url{http://dx.doi.org/10.1016/j.pbiomolbio.2016.08.004}
}

\epigraph{Theory and fact are equally strong and utterly interdependent; one has no meaning without the other. We need
theory to organize and interpret facts, even to know what we can or might observe. And we need facts to validate
theories and give them substance. Theory and fact are equally strong and utterly interdependent; one has no
meaning without the other. We need theory to organize and interpret facts, even to know what we can or might observe.
And we need facts to validate theories and give them substance.}{S.J. Gould (1998) }

\section{ Introduction}
Scientific theories provide organizing principles and construct objectivity by framing observations and experiments
(Longo \& Soto, this issue).  On the one hand, theories construct the proper observables and on the other they provide
the framework for studying them. Good theories, like Newton’s law of inertia and the conservation of momentum from his
3\textsuperscript{rd} law were never abandoned but were reinstated while physics underwent further theoretical changes.
Indeed, a deeper understanding of these principles was gained through E. Noether’s theorems which justify the
conservation properties of energy and momenta in terms of symmetries in the state equations (van Fraassen 1989).
However, a theory does not need to be “right” to guide the praxis of good experiments. Even a ``wrong'' theory can be
useful if, when proven incorrect it is modified or even dismissed. 

Here we demonstrate that the application of the principles we propose to use for the construction of a theory of
organisms results in a better understanding of morphogenesis (the generation of biological form) than the common
practice of using metaphors derived from the mathematical theory of information as theoretical background [for a
critique see Perret and Longo in this issue, and (Longo et al. 2015)]. Our approach diverges from the biophysical
methodology which is based on conservation principles and their associated symmetries, on the one hand, and with
optimization principles, on the other. In contrast, biology is about an incessant breaking of symmetries (Longo et al.
2015; Longo and Montévil 2011; Longo and Soto, this issue). Taking mammary gland morphogenesis as an example, here we
show that our theoretical principles are useful to provide a framework for the mathematical modelling of tissue
morphogenesis. We will focus on the manner in which we propose to deal with cellular behavior.

\section{Theoretical principles}
We begin by identifying a  foundational principle, the default state of cells, which is proliferation with variation and
motility. This default state is a manifestation of the agency of living objects, and thus, a cause; it does not need an
explanation or an external cause (Longo et al. 2015, Soto et al. 2016). The default state is what happens when nothing is done to the
system. Second, we adopt the notion that organismal constraints prevent the expression of the default state. This means
that constraints determine when proliferation with variation and/or motility are allowed to instantiate. Third, we
consider it essential to stress that biological processes make full sense only in the context of the organism in which
they take place. As stated by Claude Bernard: “The physiologist and the physician must never forget that the living
being comprises an organism and an individuality. If we decompose the living organism into its various parts, it is
only for the sake of experimental analysis, not for them to be understood separately. Indeed, when we wish to ascribe
to a physiological quality its value and true significance, we must always refer to this whole and draw our final
conclusions only in relation to its effects in the whole.” We address this aspect of biological integration using the
notion of closure of constraints (see Mossio et al this issue). These constraints are considered “local invariants”
because they do not change at the time scale of the process they influence. In an organism, these constraints depend
collectively on each other thus attaining closure. In turn, closure provides an understanding of the relative stability
of biological organizations. Fourth, organisms spontaneously undergo variation.  A fundamental generator of variation
is the default state. Unlike physical systems, biological ones are not framed by invariants and invariant preserving
transformations. Instead, the flow of time is associated with qualitative changes of organization that cannot be stated
\textit{a priori}. This original feature is directly related to the historical nature of biological objects, since a
specific object is the result of this unpredictable accumulation of changes (Longo et al. 2015, this issue, Montévil et
al, this issue). Another notion that becomes fundamental through this principle of variation is that of contextuality. Indeed,
understanding biological organization requires taking into account its interaction with the surrounding environment,
both at a given time-point and through the successive environments that biological objects traverse (Soto and
Sonnenschein 2005)(see Miquel and Hwang, Montévil et al and Sonnenschein and Soto  in this issue). In addition to the
role that constraints play in canalizing processes, they make possible the appearance of new constraints and thus
changes of organization. Lastly, the framing principle states that biological phenomena should be understood as the
non-identical iterations of morphogenetic processes. Biological processes iterate at all levels of organization.
Organization involves iteration through the circularity of closures, but organization itself is also iterated as
reproduction. Inside organisms, structures are also iterated, for example in the case of branching morphogenesis. In
all cases, the principle of variation applies so that each iteration may be associated with unpredictable qualitative
changes (Longo et al. 2015).

Together these principles provide a genuinely biological framework for the understanding of organismal phenomena. This
framework combines the integrative viewpoint inherited from physiology, the centrality of biological variation that
derives from the theory of evolution and the default state that links organismal and evolutionary biology. 

\section{ The mammary gland as a model system}
The mammary gland is made up of two main tissue types, namely, i) the epithelial parenchyma, its function is to produce
and deliver milk,  and ii) the stroma which surrounds and supports the epithelium. The stroma is composed of various
cell types (fibroblasts, adipocytes, and immune cells), blood vessels, nerves, and an extracellular fibrous matrix of
which the main component is collagen type-I. In the resting gland the epithelium is organized into a ductal tree.
During pregnancy a second epithelial compartment, the alveoli, develop from the ducts; these are the structures that
produce and secrete milk. Throughout development, reciprocal interactions between the epithelium and the stroma are
responsible for the structure and function of mammary glands. Perturbations of epithelial-stromal interactions result
in various pathologies including neoplasms (Soto and Sonnenschein 2011, Sonnenschein and Soto 2016).

The mammary gland undergoes morphological and functional changes throughout life. Mammary organogenesis has been studied
in most detail in rodents. In mice, the mammary placodes become visible between embryonic day (E) 11 and 12 and they
then develop into mammary buds by E13. At this time, several layers of mesenchyme condense surrounding the buds in a
concentric fashion. In female mouse embryos the mammary bud sprouts and invades the presumptive fat pad. At E18, the
mammary epithelium consists of an incipient ductal tree (Balinsky 1950; Robinson et al. 1999). Although fetal mammary
gland development occurs even in the absence of receptors for mammotropic hormones suggesting that these hormones are
not required at this stage, fetal mammary morphogenesis can be altered by exposure to hormonally-active chemicals
(Vandenberg et al. 2007; Vandenberg et al. 2007). From the onset of puberty, the development of the mammary gland is
subject to hormonal regulation. At the onset of puberty, estrogens induce the formation of club-shaped structures at
the end of the ducts, called terminal end buds (TEBs). Thereafter, the epithelium begins to fill the fat pad and
branches. Progesterone induces lateral branching. If pregnancy occurs, prolactin in combination with estrogen and
progesterone initiates a characteristic lobuloalveolar development [reviewed in (Brisken and O'Malley 2010)]. When
lactation ceases, involution of the alveolar structures occurs and the mammary gland returns to its resting state. 

\subsection{Biological models for the study of mammary gland biology}

3D culture systems allow for the dynamic study of epithelial morphogenesis and the organization of the stroma. These
models are intended to mimic conditions prevailing in a living organism while reducing the number of constraints
present \textit{in vivo} to those which theoretically and/or empirically are considered to be the most relevant ones
for the subject study. When designing a 3D culture model one must first define the main characteristics of the target
tissue that the model aims to reproduce, and which stage of mammary gland development the investigator is interested in
reproducing \textit{in vitro}. The objective of mimicking the tissue of origin is tempered by the need to make it
manageable by reducing the model to a few components. This allows the researcher to infer from these results the
contribution of these components to the mammary gland phenotype \textit{in vivo}. The resulting model may then be
compared to more complex ones resulting from the step-wise addition of relevant components, and eventually to the
behavior of the gland \textit{in situ}.  

Epithelial–stromal interactions can be studied using 3D co-cultures of epithelial and stromal cells by analyzing matrix
remodeling and epithelial morphogenesis. With regard to matrices the most biologically relevant ones are those that
provide the structure and rigidity of the model tissue that allows the cellular components to attain characteristics
seen in the breast. 

Here we focus on collagen-based matrices since collagen is a main component of the mammary stroma that allows for breast
epithelial cells to organize into structures that closely resemble those observed \textit{in vivo} (Krause et al.
2012; Krause et al. 2008; Dhimolea et al. 2010; Speroni et al. 2014; Barnes et al. 2014). 

\subsubsection{ MCF10A 3D culture model}
To test the mathematical model, we used data generated using the MCF10A 3D culture model described in (Barnes et al.
2014). In this 3D culture model, the proportion of acinar and ductal structures can be modified by changing the
concentration of reconstituted basement membrane (Matrigel) in the extracellular matrix (ECM) (Krause et al.
2008; Barnes et al. 2014). Briefly, MCF10A human breast epithelial cells were seeded in either bovine type-I collagen
matrix or in mixed matrices containing collagen and Matrigel at 5 and 50\% v/v. The final collagen concentration in all
gels was 1.0 mg/ml. Gels were prepared by carefully pouring 500 $\mu $l of the cell-matrix mixture into wells of a
glass-bottomed 6 well plate. The gels were allowed to solidify for 30 min at 37°C before adding 1.5 ml of cell
maintenance culture medium into each well. Cultures were maintained at 37°C in an atmosphere containing 6\%
CO\textsubscript{2}/94\% and 100\% humidity 5 or 7 days, and the medium was changed every 2 days. Live cell-imaging was
conducted with a Leica SP5 microscope (Leica-Microsystems, Germany). Simultaneous reflectance confocal microscopy (RCM)
and bright-field images were acquired using a 40×, 1.1 numerical aperture, water immersion objective with separate
photomultiplier tubes using the Argon 488 nm laser line. The RCM signal was collected between wavelengths of 478–498 nm
with a pinhole size of 57 $\mu $m. For additional details see  (Barnes et al. 2014).

\subsection{ Current mathematical models of mammary gland morphogenesis}
Several types of mathematical models have been used for the study of mammary gland morphogenesis. These models focus on
distinct aspects of this phenomenon, and address a given aspect at a particular time and space scale.  These diverse
models also focus on different kinds of determinants, some chemical, some mechanical, some both.  
\begin{table*}[!ht]
\begin{small}
 \begin{tabular}{|m{1.1in}|m{1.4in}|m{1.4in}|m{1.4in}|m{0.8in}|}
\hline
{\bfseries Object studied} &
{\bfseries Modeling cell proliferation} &
{\bfseries Modeling motility} &
{\bfseries Implicit default state} &
{\bfseries Reference}\\\hline
{\bfseries Collagen network remodeling} &
Not discussed. &
ECM constrains movement. &
Motility. &
(Harjanto and Zaman 2013)\\\hline
{\bfseries Collagen and fibroblasts} &
Not discussed. &
Cells spontaneously exert forces and move. Stress may prevent motion. &
Motility. &
(Dallon et al. 2014)\\\hline
{\bfseries Acinus in 2D} &
Limited space prevents proliferation. Although not included in the math model, it is stated that proliferation is
regulated by signals/growth factors.  &
Not discussed. &
Proliferation (quiescence is invoked the discussion). &
(Rejniak and Anderson 2008)\\\hline
{\bfseries Acinus in 3D} &
Only cells adjacent to the basement membrane proliferate. Cells have a proliferative potential that decreases at each
division. &
Not discussed. &
Shifts from proliferation to quiescence as time elapses. Not discussed for inner cells. &
(Tang et al. 2011)\\\hline
{\bfseries Review on biophysical cell self-assembly} &
No proliferation. &
Cells show trend to move modeled by a parameter formally similar to temperature. &
If this {\textquotedbl}temperature parameter{\textquotedbl} is an intrinsic property of cells, the default state is
motility, otherwise quiescence. &
(Neagu 2006)\\\hline
{\bfseries Terminal End Buds} &
\multicolumn{2}{m{2.6789598in}|}{Not causal analysis (measured or assessed indirectly).} &
NA &
(Paine et al. 2016)\\\hline
{\bfseries Epithelial tree} &
Components resulting from the action of matrix metalloproteinases are inferred to stimulate cell proliferation. &
Not discussed. &
Quiescence. &
(Grant et al. 2004)\\\hline
\end{tabular}\end{small}
 \caption{ Cell behavior according to current mathematical models of mammary gland morphogenesis. }
 \label{table1}
\end{table*}

Oftentimes the modelers make mathematical hypotheses on the behavior of cells without making explicit the broader
biological significance of these hypotheses. Our mathematical model instead is based on general biological principles.
Some models seem to implicitly rely on the default state that we propose: that is, cells spontaneously move or
proliferate and the model discusses specific constraints on these behaviors. In other models, cells are quiescent
without any constraints acting on them and chemicals stimulate proliferation and/or movement without removing
constraints. Still, in other models even opposite behaviors co-exist and no explicit attempt to reconcile these
opposites is made. In Table \ref{table1} we review the way several models deal with cell behavior.

In these models assumptions on cell behavior are largely \textit{ad hoc}, varying  from one model to another.
Occasionally, the model and its interpretation in the Discussion are inconsistent. For example, in (Rejniak and
Anderson 2008) proliferation is constrained by the available space but it is also attributed to
{\textquotedbl}signals{\textquotedbl}.

Various models implicitly adopt the premise that the default state of cells is proliferation or motility. Other models
(Grant et al. 2004) do not adopt this premise; however they may be reinterpreted by adopting our principles. In the
Grant et al model, the matrix metalloproteinases are assumed to have a positive effect on proliferation by degrading
the ECM. Because the authors also infer that the default state is quiescence, they assume that this degradation
produces a chemical which stimulates proliferation. A simpler hypothesis that we favor is that this degradation removes
the mechanical constraint of the ECM on the default state of cell proliferation with variation and motility. To our
knowledge, no model of mammary gland morphogenesis has taken both aspects of our default state into account. 

The main aim of this article is to emphasize that making explicit the assumption that the default state is proliferation
with variation and motility enables us to model morphogenesis on precise theoretical bases. The models reviewed above
focus on different aspects of mammary gland morphogenesis. For instance, models of ECM remodeling focus on fibroblasts
and do not discuss epithelial morphogenesis. Several agent-based models focus on the behavior of epithelial cells
during acinus formation in conditions that preclude ductal morphogenesis.  Finally, other models focus on larger scale
organogenesis but do not provide a detailed account of cellular behavior. Determinants of these models are either
chemical, mechanical forces or empty space. Note however that in most cases morphogenesis is assumed to be driven by
chemical interactions (Iber and Menshykau 2013).  This article will focus on the formation of epithelial structures,
mostly ducts but also acini on the basis of mechanical interactions between cells and between cells and the ECM.  

Mammary gland morphogenesis \textit{in vivo} requires an interplay between the epithelial compartment and the stroma
which contains fibroblasts, adipocytes and ECM.  Our simplified model contains epithelial cells and a stroma devoid of
cells but containing the ECM which \textit{in vivo} would be a product of the stromal cells. The assumption that the
ECM is the mechanically important component of the stroma opens the possibility to explore a set of minimal and
manageable conditions that allow for a bi-stable determination of acini and ducts, the two structures that characterize
the mammary gland parenchyma. In the rest of the text we will discuss all experimental results and mathematical
assumptions under the hypothesis that the default state of cells is proliferation with variation and motility.

\subsection{ From the 3D culture model to a mathematical
model}
\subsubsection{ Proliferation}
As expected from the default state, breast epithelial cells proliferate maximally in serumless medium. Addition of
hormone-free serum to culture medium results in a dose-dependent inhibition of cell proliferation. This is due to the
effect of serum albumin, a main constraint for the proliferation of estrogen-target cells. This constraint could be
lifted by lowering albumin or serum concentration, a procedure easily performed in 2D cell cultures, or by adding
estrogens, which is the natural way the organism uses to neutralize the effect of serum albumin (Sonnenschein et al.
1996).

Additional constraints are those imposed by cell-cell contact, which are weaker in 2D culture than in 3D. This is
because 3D structures allow for cell-cell contact in practically all directions, while in 2D these contacts are more
restricted. For example, the estrogen-sensitive T47D cell line is inhibited from proliferating when placed in medium
containing serum. This constraint is lifted by the addition of estrogen.  When placed in 2D culture, the cell number
ratio between serum plus estrogen and serum without estrogen is 3.75, while the same experiment performed in 3D culture
results in a ratio of 2.5 (Speroni et al. 2014). These data indicate that the organization of cells into epithelial
structures constrains cell proliferation more effectively than the one in 2D culture, where cells are not organized into
closely packed epithelial structures. Additionally, in 2D cultures cells attach to the bottom of the culture dish, a
surface that is exceedingly more rigid than the conditions these very cells encounter within the tissue of origin. In
contrast, 3D cultures could be engineered to mimic the rigidity of the tissue of origin.

\subsubsection{ Motility}
In classical mechanics, motion is a consequence of external forces. In biology, the situation is different but
compatible with mechanics. Cells need a configuration of forces to be able to move; for example, they need a support to
be able to crawl on it, or fibers to which they can attach and pull in order to move. However, unlike inert objects,
cells initiate movement on their own. In other words, they are autonomous agents which express their default state (Sonnenschein and Soto, 1999).
Cells move unless there are constraints which prevent them from doing so. Reciprocally, a given mechanical force acting
upon biological entities such as cells produces clearly different effects than forces acting on inert matter (Soto et
al. 2008; Longo and Montévil 2014). For example, gravity in mechanics is just a force proportional to the mass of the
object and oriented towards the center of the earth. In biology, however, gravity becomes a constant constraint, which
has not been altered since the origin of life.  Biological organization reacts to it in various ways. For example, swim
bladders, wings, limbs and tree trunks are responsive to gravitational force but are not explained by it. The behavior
of molecules in cells and the overall behavior of tissues and organs are massively impacted  when in microgravity
conditions (Bizzarri et al. 2014).

\begin{figure*}
\centering
 \includegraphics[scale=0.2]{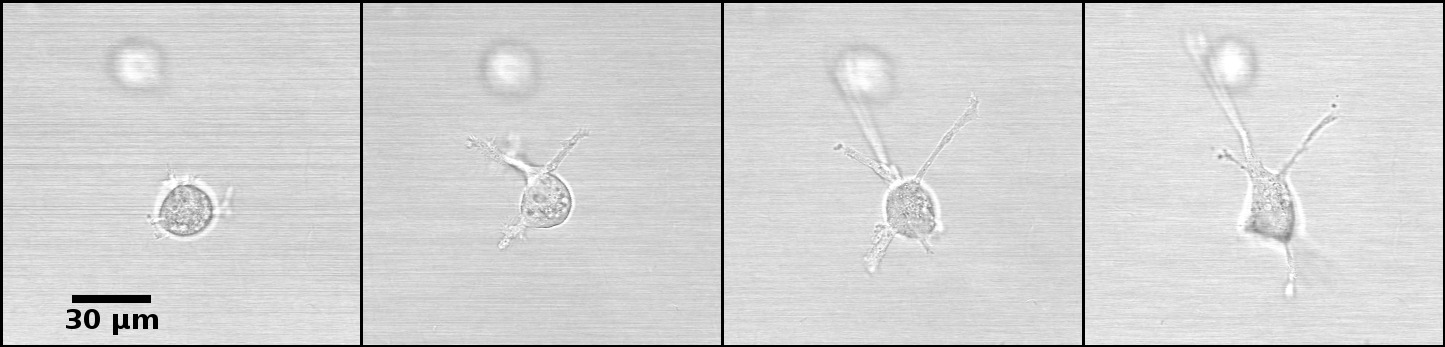} 
\caption{\textit{Projections of breast epithelial cells seeded in a fibrilar matrix.}  Soon after seeding cells emit projections
in all directions; these projections are involved in collagen organization. Still images from video \ref{vid1} at 3, 4, 5 and 6
hours after seeding.
}\label{fig1}

\end{figure*}
\begin{figure}
\centering
 \includegraphics[scale=0.7]{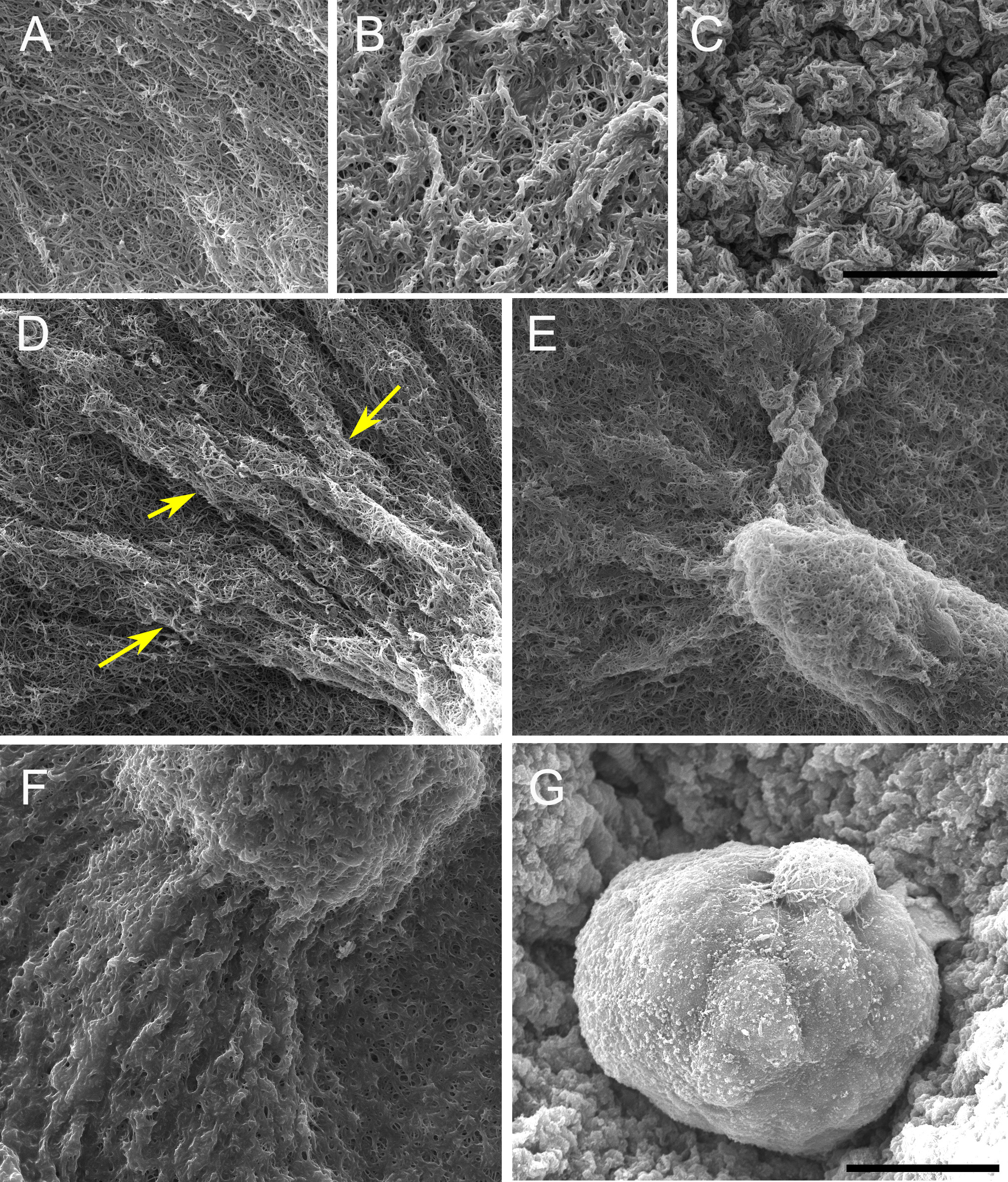} 
\caption{\textit{SEM images of epithelial structures and their matrix at day 5.} (A) Collagen fibers are clearly distinguished
in a collagen-only matrix. (B) Addition of 5\% Matrigel results in a globular rather than fibrilar matrix. (C) The
globular matrix is more compact in 50\% Matrigel. (D) Tube-like cell processes (arrows) are observed at the tip of a
duct in a collagen-only matrix. (E) Ductal and (F) acinar structures in 5\% Matrigel; Matrigel forms a localized
coating in areas surrounding the acinus. (G) An acinus grown in 50\% Matrigel; collagen fibers are not visible. Scale
bar, 10 $\mu $m in A to C and 15 $\mu $m in D to G.  Reproduced with permission from (Barnes et al. 2014). 
}\label{fig2}

\end{figure}

\begin{figure}
\centering
 \includegraphics[scale=0.015]{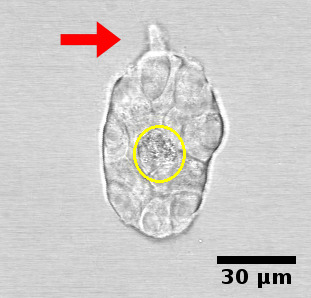} 
\caption{\textit{Breast epithelial cells forming an acinus in a non-fibrilar matrix at day 4.} Cells display limited motility
and emit only short projections into the matrix (arrow). Cells rotate and divide resulting in the formation of an
acinus, a sphere with a central lumen (circle). Still image from video \ref{vid2}. 
}\label{fig3}

\end{figure}

The constraints to motility that cells experience in the tissue environment can be modeled in a 3D culture system. In
this system, the tissue environment is recreated by the matrix in which the cells are seeded. Cells exert their
motility by using filopodia and pseudopodia. When they encounter a structure such as a fiber they can use it for
locomotion or pull on it to attach. Breast epithelial cells seeded in a fibrilar matrix emit projections in all
directions soon after seeding (Figure \ref{fig1}; video \ref{vid1}). This process allows cell elongation which precedes the formation of
ducts (tubular structures) and branching (Barnes et al. 2014). Amorphous, non-fibrilar matrix proteins as well as
fibers also oppose migration and cause resistance due to their relative rigidity and their lack of pores (Figure \ref{fig2}).
Breast epithelial cells growing in a non-fibrilar matrix display limited motility and emit short projections into the
matrix that retract soon afterwards. Cells rotate and divide resulting in the formation of an acinus, a sphere with a
central lumen (Tanner et al. 2012) (Figure \ref{fig3}; video \ref{vid2}). Cell movement is also constrained by the pore size of the
matrix, this is mostly determined by fiber alignment, fiber density and abundance of non-fibrous matrix materials.
Other factors such as pore size and matrix rigidity could contribute to the morphological differences of the epithelial
structures. Pore size is bigger in the fibrilar matrix than in the globular matrix (Figure \ref{fig2}). The globular matrix is
stiffer than the fibrilar matrix, however this proved to be a minor contributor to epithelial phenotype compared to
collagen fiber distribution (Barnes et al. 2014). 

\paragraph{ Cell adhesion }

\begin{figure*}
\centering
 \includegraphics[scale=0.35]{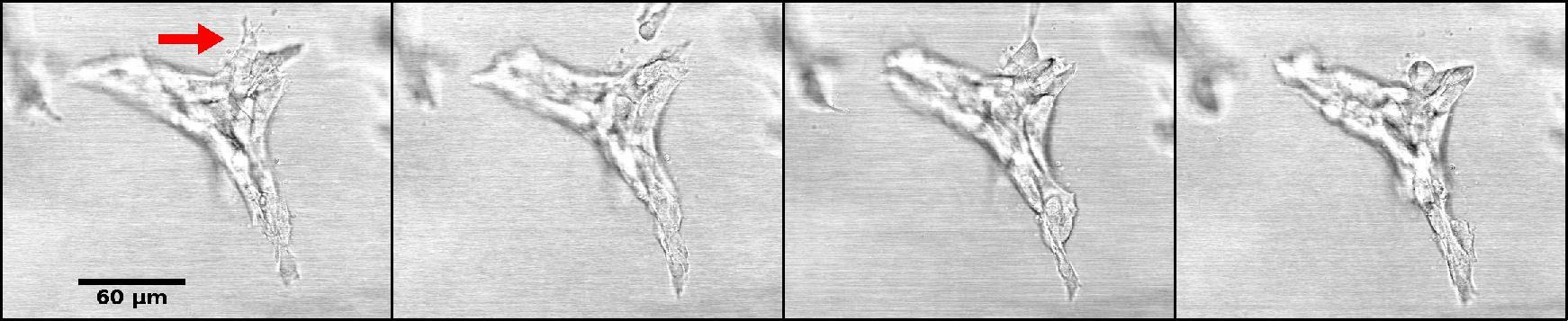} 
\caption{\textit{Branching duct at day 7 of culture.} A cell (arrow) detaches from the main structure and is incorporated back
into the same structure. Still images from video \ref{vid3}, each frame corresponds to one hour.
}\label{fig4}
\end{figure*}

Cells can adhere to each other after division of their progenitor; they can also attach to any migrant cells they
encounter. Once an epithelial structure, such as a duct or an acinus, is formed, adhesion will be maintained as new
cells are formed or replaced. Adhesion constrains the motion of cells. During ductal morphogenesis, single cells can
detach from the main structure and may also be incorporated back into the structure from which they detached (Figure \ref{fig4},
video \ref{vid3}). This phenomenon suggests an environment-sensing strategy as well as a means used by epithelial structures to
modify the matrix prior to growth in a certain direction. During lumen formation, cells migrate toward the periphery of
the epithelial structure leaving a space that will be filled by fluid.

\subsubsection{ Determination of the system}
\begin{figure}
\centering
 \includegraphics[scale=0.2]{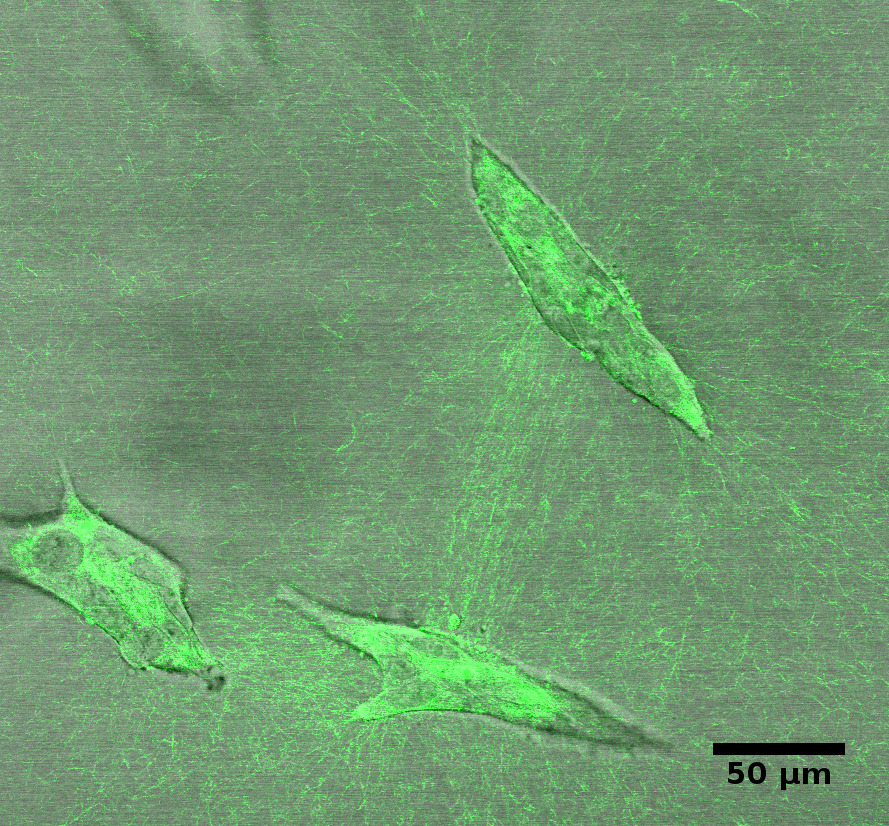} 
\caption{\textit{Collagen fibers and breast epithelial structures after 6 days in culture.} Cells organize collagen in a
collagen only matrix and the collagen bundles (green) facilitate the merging of epithelial structures. Still image from
video \ref{vid4}. (For interpretation of the references to colour in this figure legend, the reader is referred to the web version of this article).
}\label{fig5}

\end{figure}

The increase in cell number due to the unconstrained default state brings about re-distribution of fluids,
reorganization of fibers and a certain degree of matrix compression, and/or matrix degradation. Elongation is
accompanied by fiber organization into bundles projecting in the direction of the ductal tip (Barnes et al. 2014).
Collagen bundles facilitate the merging of epithelial structures initially positioned at a long distance range (Guo et
al. 2012)(Figure \ref{fig5}, Video \ref{vid4}).

\begin{figure}
\centering
 \includegraphics[scale=0.4]{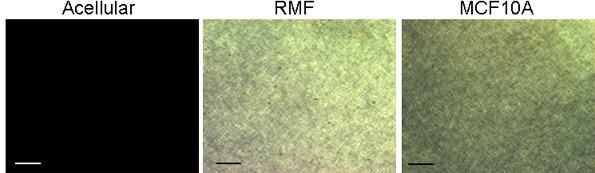} 
\caption{\textit{Evidence of cell activity on collagen organization.} Collagen fibers rearrangement in a fibrillar matrix
containing no cells, fibroblasts (RMF) or MCF10A breast epithelial cells, 24 h after seeding. Whole mount picrosirius
red staining/polarized light imaging; scale bar 100 $\mu $m. Reproduced with permission from (Dhimolea et al. 2010).
}\label{fig6}

\end{figure}

Acellular collagen gels contain small fibers that are not seen using the classical picrosirius red/polarized light
method to visualize collagen fibers. However, a similar collagen gel containing cells reveals a very different picture;
twenty-four hours after seeding the cells, small yellow fibers are detected (Figure \ref{fig6}). This experiment shows that
cells organize collagen fibers. In other words, cells exert forces upon fibers, and fibers transmit these forces for
quite long distances (Guo et al. 2012).

\begin{figure}[!ht]
\centering
 \includegraphics[scale=0.75]{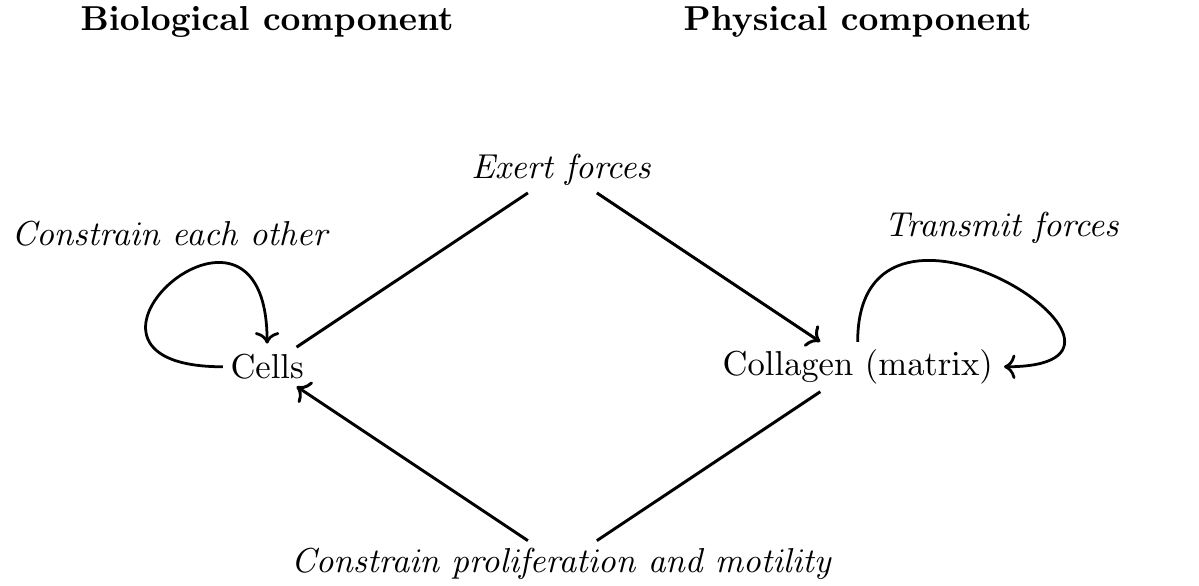} 
\caption{\textit{Schematic representation of the determination of the 3D biological model.} We analyze morphogenesis as the
interaction between a physical and a biological component. The physical component is mostly made out of collagen and is
determined by physical principles in the absence of cells. Collagen undergoes spontaneous gel formation; however, its
structure does not change much spontaneously after this process. Cells are agents which act upon other cells and
collagen fibers organizing them. The formation of epithelial structures by cells is modeled from the default state;
cells and collagen constrain this default state.
}\label{fig7}

\end{figure}

As collagen fibers progressively organize, they constrain the proliferation and motility of cells. These constraints may
be positive like those that facilitate cell migration along fibers or negative like the ones hindering migration
orthogonally, and those due to “pore” formation when the collagen fibers organize into a network. Additionally, cells
constrain other cells mechanically. The reciprocal interactions between the collagen and the cells are illustrated in
Figure \ref{fig7} which emphasizes the separation of the system into a physical component, the collagen, and a biological
component, the cells. In this diagram the behavior of the cells is determined by the default state and the constraints
exerted on it.

\section{ Implementing a model based on these principles}
We propose a mathematical model of morphogenesis in 3D cultures that we analyze by computer simulations. In this article
this model is used  as a proof of concept. A mathematical description of our model is provided in Appendix \ref{appendix}. Our model
is based on the principles that we propose for the construction of a theory of organisms. Our theoretical framework
restricts what is acceptable in order to model cellular behavior. For example, it is  unacceptable for cells to be
proliferatively quiescent without an explicit constraint keeping them in this state. Of course, in the presence of a
strong constraint cells will become quiescent and will remain as such for the duration of the constraint. 

\subsection{Components of the model}

The 3D culture gel is represented as a lattice, in three dimensions. The model is based on different layers that
interact with each other:

\begin{enumerate}[i)]
\item A mechanical forces layer. Each elementary cube of our model exerts forces on the adjacent cubes. Forces propagate
in space. Their orientation and propagation depends on the orientation of collagen fibers.
\item A cellular layer, which has three possible states: presence of a live cell, a dead cell (in the case of lumen
formation), or no cell, in which case the cube contains ECM. 
\item The collagen layer is approached at a mesoscopic scale and does not represent individual fibers. Each cube of
collagen has a main orientation which will transmit forces farther, and this orientation is random in the initial
conditions. This orientation is also relevant for cellular behavior (see below). Note that this layer represents the
cytoskeleton in the cubes occupied by cells. Collagen tends to align with forces that are exerted on it.
\item An inhibitory layer. Cells around the lumen produce a short range inhibition of both proliferation and motility.
This layer can be interpreted either as the formation of a basal membrane or as a chemical inhibitor. Both aspects are
biologically relevant and could be split into two different types of inhibition in future work. 
\item A “nutrient{\textquotedbl} layer governed by diffusion. Cells consume nutrients. If there are not enough
nutrients, cells die leading to lumen formation. The main point here is not to produce an accurate account of lumen
formation, but instead to take into account the consequences of lumen formation on morphogenesis as a lumen would
disrupt the transmission of forces.
\end{enumerate}
\subsection{ Cellular behavior:}

Cells express four different behaviors:

\begin{enumerate}[i)]
\item Cells exert forces on each other and on the collagen network. The orientation of these forces is influenced by
both the neighboring cells and there is also a random component. The magnitude of the force depends on the content of
the neighboring cube. More precisely, the force exerted depends on the position with respect to the cell, the
orientation of the cytoskeleton of the cell and the orientation of the collagen fibers in the case of an action upon a
collagen fiber or fascicle. Additionally, cells tend to oppose strong mechanical stress.
\item Cells have a defined generation time and divide, except when there are constraints which prevents them from doing
so. The new cell will occupy a random spot. When this spot is already occupied, the cell cannot proliferate; the cell
will make another attempt at the next iteration of the simulation loop.
\item Cells move randomly unless this movement is constrained.
\item Cells die when they lack nutrients. Unlike the other behaviors, this one is an \textit{ad-hoc} addition to create
a lumen, since the  steps involved in lumen formation are not well known. There is evidence for cell death
and for cell migration, but the cue provoking lumen formation is unknown.
\end{enumerate}
 In this analysis, cell motility has two components: the forces exerted by cells and cell movement. Cell movement involves
detachment and reattachment to other cells and to the extracellular matrix.

\subsection{ Constraints on proliferation:}

\begin{enumerate}[i)]
\item Cells tend to proliferate along the direction of forces. The stronger the forces, the stronger the constraint
becomes. In the case of three cells being aligned, the one in the middle may be unable to proliferate if there is a
force in the direction of their alignment.
\item Strong mechanical stress slows down proliferation; this constraint is not required for the model to work.
\item The inhibitory layer prevents proliferation.
\end{enumerate}
\subsection{ Constraints on cell movement: }

\begin{enumerate}[i)]
\item A strong mechanical stress slows down movement; this is not required for the model to work. Also, as the number of
neighboring cells increases, the stronger the effect of cell adhesion, which prevents the cell from initiating
movement.
\item Movement is facilitated when it occurs along collagen fibers.
\item Movement is facilitated towards other cells.
\item The inhibitory layer prevents movement.
\end{enumerate}
\subsection{ Results of implementing the mathematical model}

\begin{figure*}[!ht]
\centering
 \includegraphics[scale=0.9]{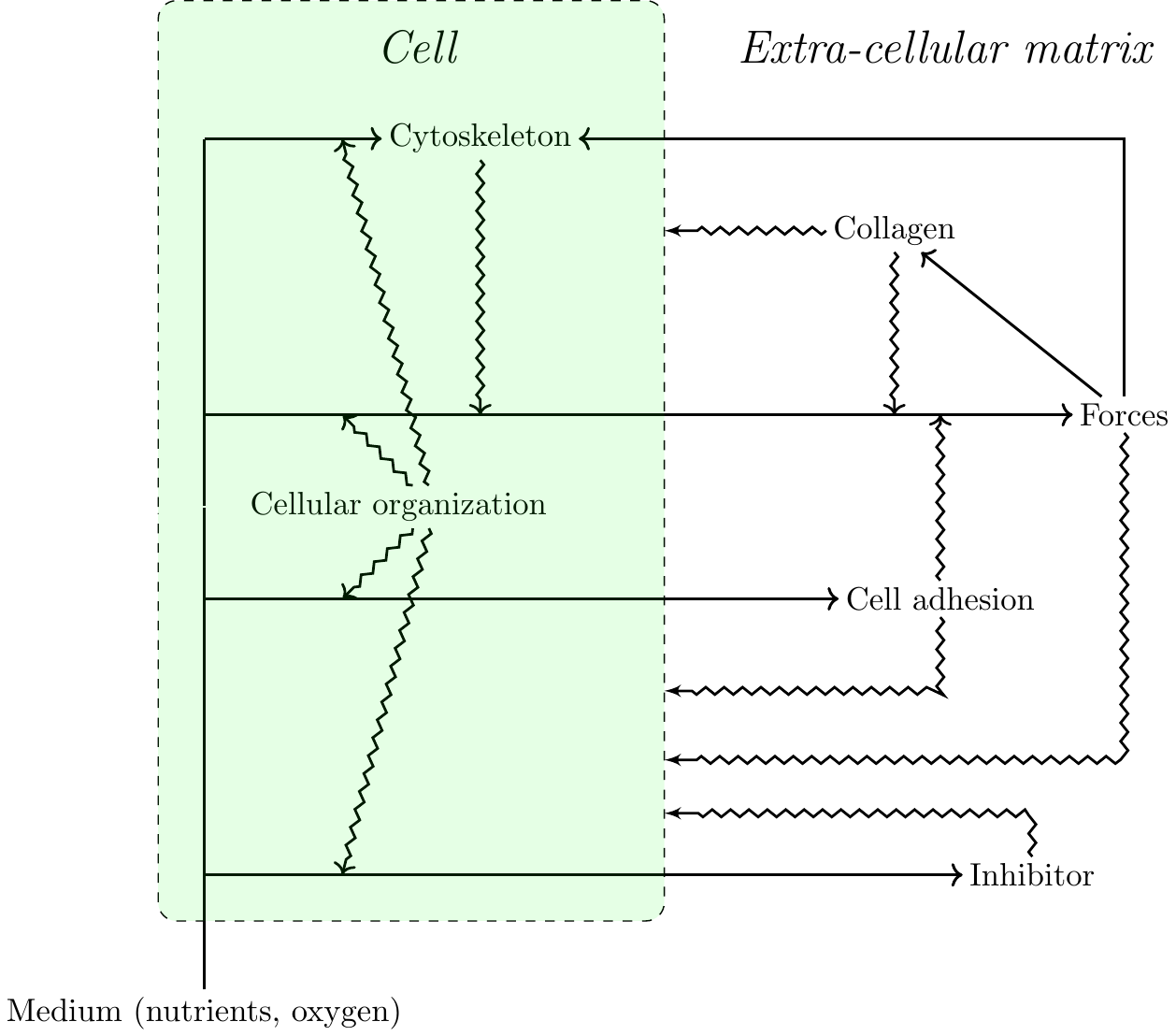} 
\caption{\textit{Schematic representation of the theoretical model.} Chemical energy and chemicals from the medium are
transformed by cells into the constraints at play in the system. Processes of transformation are represented by
straight arrows.  The action of constraints on these processes are represented by zig-zag arrows. Constraints on the
default state point to a cell. All cells in the model are represented by a single cell (green). Note that we single out
the cytoskeleton as a particularly relevant aspect of cellular organization in our model. This scheme shows the
circularity of the reciprocal interactions and how cells collectively constitute their own constraints leading to
morphogenesis. 
}\label{fig8}

\end{figure*}

The model that we propose exhibits a circularity that can be interpreted in terms of closure (Figure \ref{fig8}). This
circularity concerns constraints acting on i) processes such as physical forces and ii) directly on the default state.

The implementation of this mathematical model generates biologically relevant results. When in collagen, cells form
elongated structures that can be interpreted as ducts. When we remove the effects of collagen’s fibrilar structure, in
order to mimic a globular matrix (Matrigel), cells form spherical structures.

\subsubsection{ Duct formation }

\begin{figure*}[!ht]
\centering
 \includegraphics[scale=1.3]{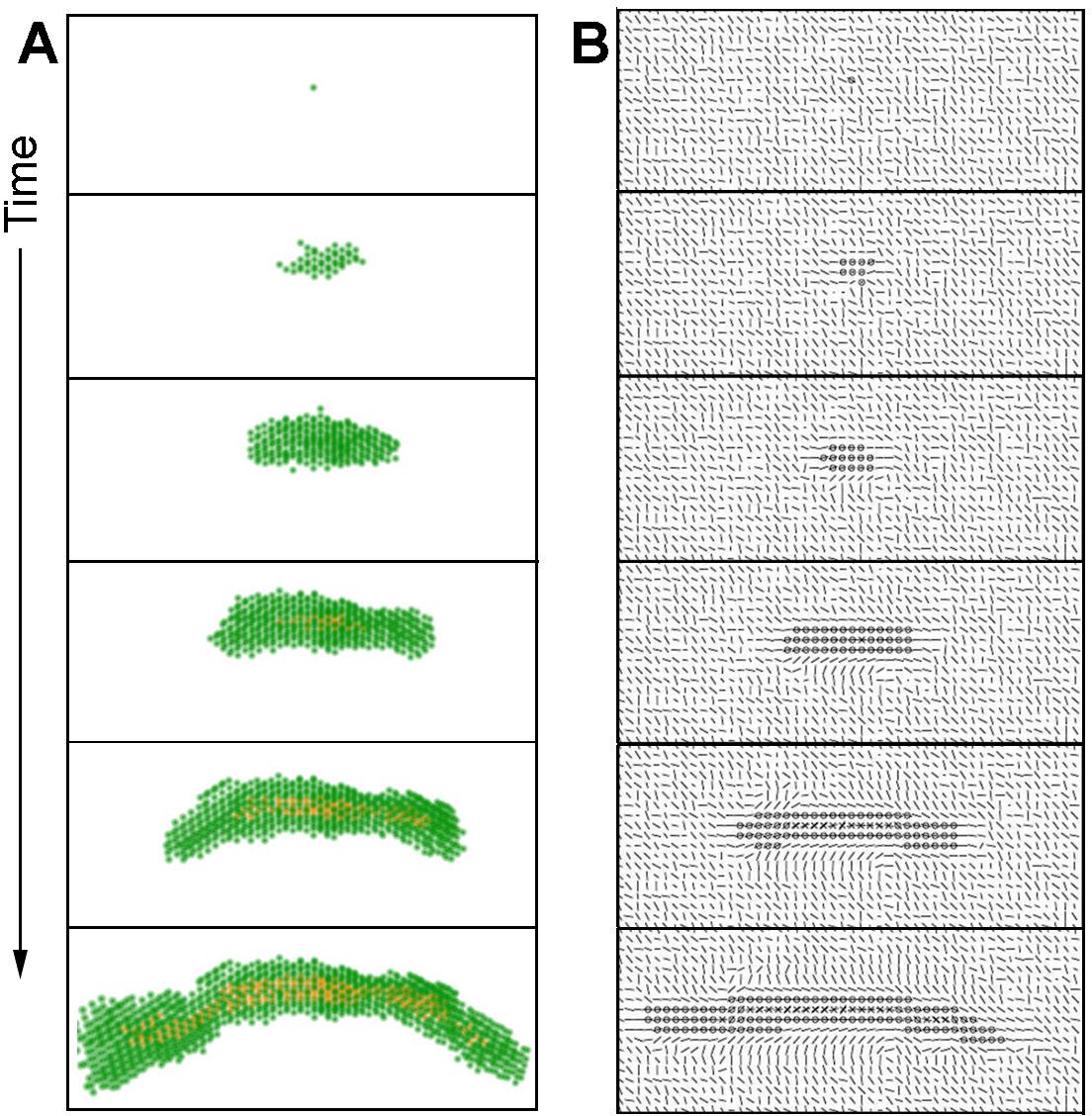} 
\caption{\textit{Formation of a duct by the mathematical model.} A) cells are represented in green and the ductal lumen in
orange, in three dimensions and over time. B) depiction of a plane of the model over time. The lines represent collagen
orientation; shorter lines mean that the orientation of collagen is mostly along the vertical axis. Circles represent
cells, and crosses represent lumen. Cells organize collagen over time by exerting forces, and collagen constrains cell
proliferation and motility. These interactions lead to the emergence of a main direction of growth.  See also
corresponding videos \ref{vid5} and \ref{vid6}.
}\label{fig9}
\end{figure*}

A single cell is surrounded by collagen. This cell starts to exert forces on the collagen and organizes it. The cell
will proliferate and may move. The epithelial structure acquires additional cells through cell proliferation. Cells
move but they mostly stay attached to the structure. Cells in the middle of the structure can neither move nor
proliferate. The cells exert forces on each other and on the collagen. This leads to the appearance of a dominant
direction in which these forces are exerted and collagen is reorganized. This direction is often that of the forces
exerted initially by the first cell and also depends on initial collagen configuration. Motility and proliferation are
facilitated along this main direction, while they are inhibited in the direction perpendicular to this force.  As a
result, the structure acquires an elongated shape. The structure grows following this dynamic until it reaches a size
large enough for the lumen to form at the thickest part of the structure, which is close to the initial position of the
first cell. In the context of lumen formation the inhibitor constrains proliferation and motility and inhibits the
growth in the width of the structure in the vicinity of a lumen. In contrast, the tips of the elongated structure are
not inhibited and elongate further without apparent restriction (Figure \ref{fig9}, videos \ref{vid5} and \ref{vid6}).

In some cases, the constraints leading to duct formation can become disorganized at one of the tips. In such case, the
main direction of forces exerted may change, leading to a change in the growth direction of the duct. In other cases,
this disorganization leads to a bifurcation in the main direction of the constraints and to branching (videos \ref{vid7} and \ref{vid8}).
These phenomena will be the subject of further studies.  

\subsubsection{Acinus formation: }

\begin{figure}[!ht]
\centering
 \includegraphics[scale=1]{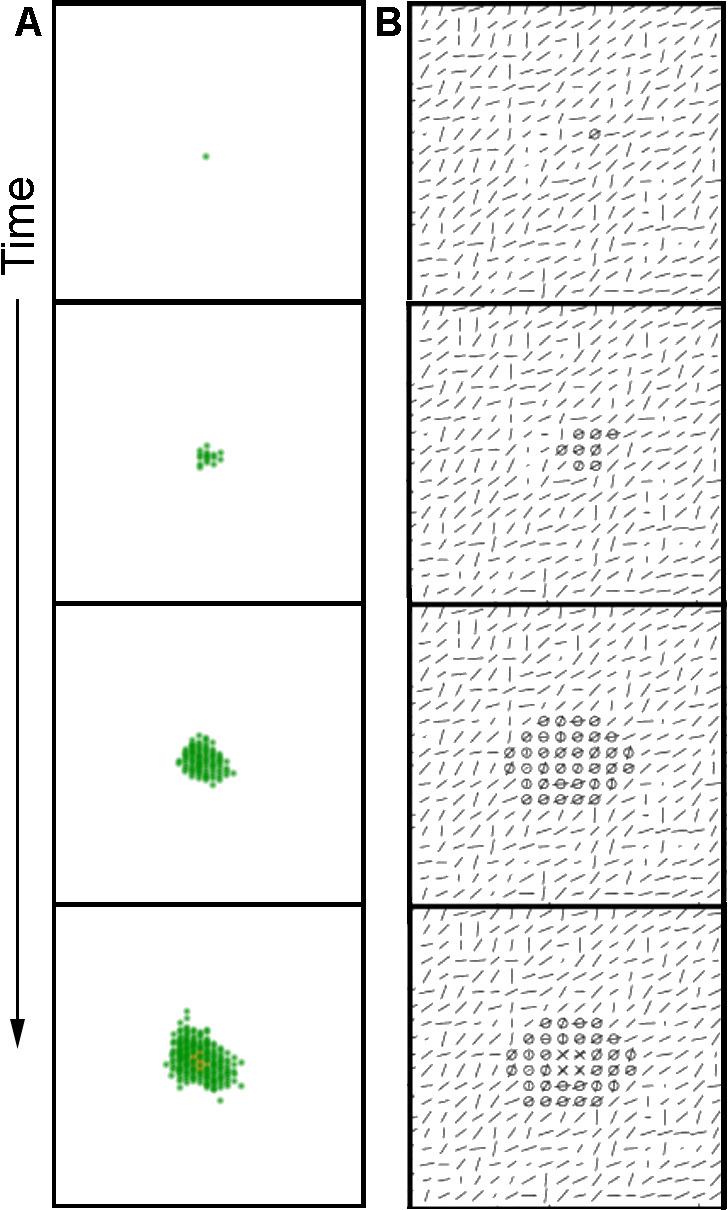} 
\caption{\textit{Formation of an acinus by the mathematical model.} A) cells are represented in green and the lumen in orange,
in 3 dimensions over time. B) depiction of a plane of the model over time. The lines represent collagen orientation;
shorter lines mean that the orientation of collagen is along the vertical axis. Circles represent cells, and crosses
represent lumen. Here we remove the interactions between the ECM and the cells in order to mimic the effect of
Matrigel.  In this condition, the cells proliferate and move in an isotropic manner leading to the formation of a
rounded structure. See also corresponding video \ref{vid9}.
}\label{fig10}

\end{figure}

The addition of Matrigel,  to a collagen matrix changes the gel properties by
coating the collagen fibers and thus hindering fiber organization (Barnes et al. 2014). In this case, collagen fibers
are not accessible to the epithelial cells, which prevents the establishment of a main direction in the forces exerted
by the cells. Moreover, these forces are exerted exclusively on  cells rather than on fibers. As a result, the
epithelial structure grows in an isotropic manner, and when the lumen is formed and the inhibitor is secreted, all
cells are constrained. The acinus being formed is a smooth structure due to the constrained motility of cells, each of
which is constrained by the many cells surrounding them (Figure \ref{fig10}, video \ref{vid9}).

\section{ The \textit{in vitro} system and the organism}
How faithfully does this in vitro system represent a phenomenon occurring inside the organism? By
accepting the reciprocal relationship between the whole (organism) and its parts, it is difficult to conceive that a
sub-system would operate in a biologically relevant fashion when outside the organism. However, there are examples in
embryology of the relative autonomy of parts at a particular point in space{}-time, as for example, the autonomy of
limb morphogenesis upon transplantation of the limb morphogenetic field to an ectopic location. Additionally,
parceling a whole does not prevent us from ascertaining whether or not the in vitro process arrives at similar outcomes
as those observed in vivo. How faithfully does this \textit{in vitro} system represent a
phenomenon occurring inside the organism? By accepting the reciprocal relationship between the whole (organism) and its
parts, it is difficult to conceive that a sub-system would operate in a biologically relevant fashion when outside the
organism. However, there are examples in embryology of the relative autonomy of parts at a particular point in
space-time, as for example, the autonomy of limb morphogenesis upon transplantation of the limb morphogenetic field to
an ectopic location.  Additionally, parceling a whole does not prevent us from ascertaining whether or not the
\textit{in vitro} process arrives at similar outcomes as those observed \textit{in
vivo}.

Which constraints are required for a relevant model of tissue morphogenesis?

\begin{figure}[!ht]
\centering
 \includegraphics[scale=0.55]{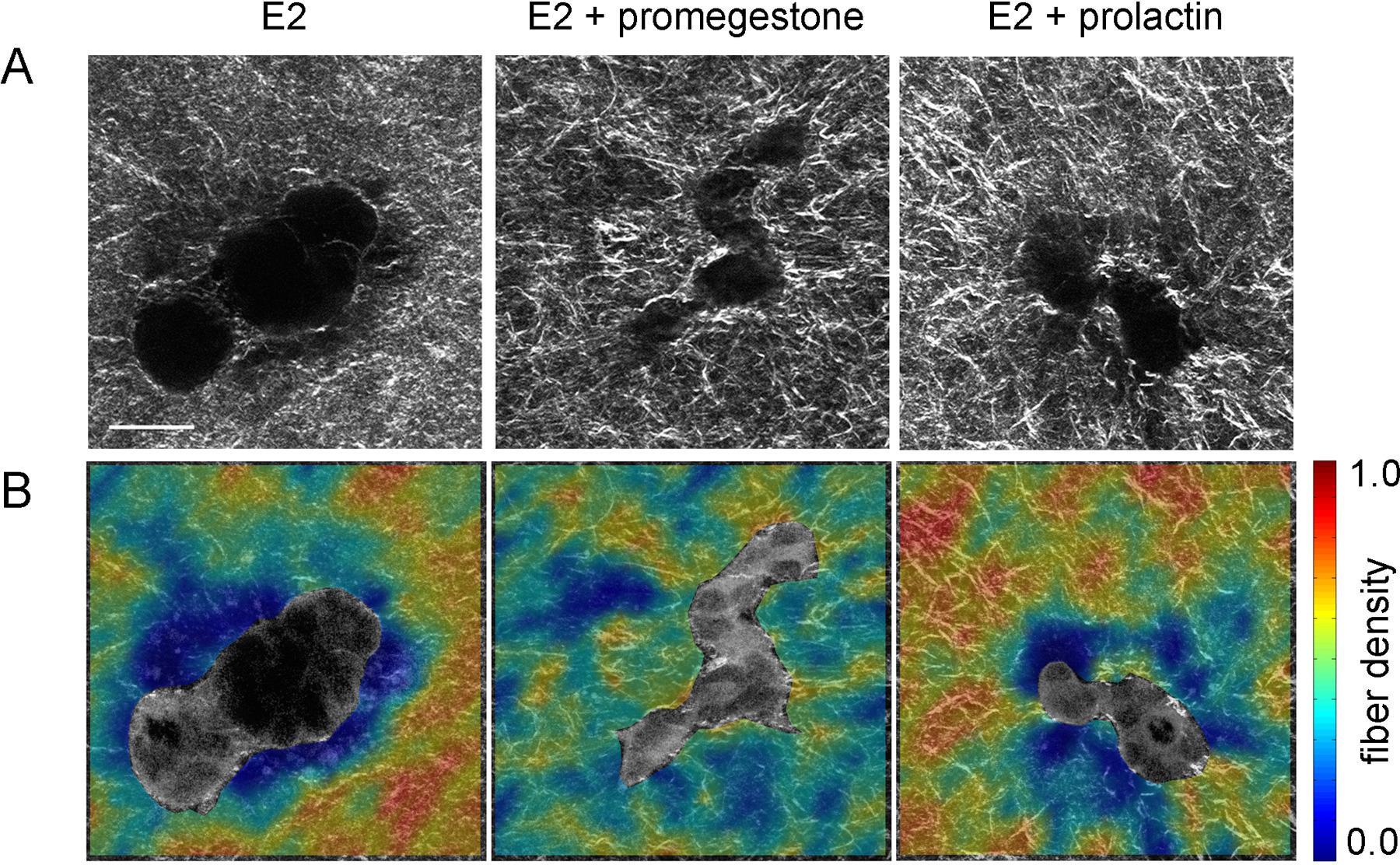} 
\caption{\textit{Effect of exposure to hormones on epithelial morphogenesis.} Hormones acting on cells generate changes in
collagen fiber organization which enable the cells to generate various epithelial organization patterns.
Hormone-sensitive epithelial structures resulting from exposure to estrogen (E2) in combination with promegestone or
prolactin were more irregular in shape than the elongated, smooth structures resulting from exposure to E2 alone.
Consistently, combined hormone treatment resulted in higher collagen density variability within 20\hspace{0.2em}$\mu $m
from the epithelial structure compared to E2 alone. (A) Second harmonic generation (SHG) images, collagen fibers in
white; (B) fiber density maps on merged SHG and two-photon excited fluorescence channels. Scale bar:
50 $\mu $m. Reproduced with permission from (Speroni et al. 2014). 
}\label{fig11}

\end{figure}

\begin{enumerate}[i)]
\item Biological meaning is construed by applying similar constraints to those which operate \textit{in }\textit{vivo}
and which seem to play a role in the determination of the phenomenon. In this way, we can “reduce” the number of
constraints to those necessary to answer our specific question. Deviations from expected results may potentially
indicate additional constraints which could then be identified.
\item Constraints that are absolutely required to allow the cells to continue being alive (pH, nutrients, temperature)
and to express their default state in conditions that replicate as much as possible the conditions present in the
organism. The “optimization” of these basal conditions is done experimentally by ascertaining that the cells can
proliferate as fast as possible and that the cellular phenotype that we wish to study is obtained. This is done in 2D
cultures. 
\item Another consideration is the historicity and specificity of biological systems.  For example, fibroblasts to be
used in a 3D culture of the mammary gland are isolated from human breast tissue and used within 6 passages to avoid
excessive deviation from the \textit{in situ} condition. Along the same lines, established epithelial cells are
maintained in standardized conditions that result in the reproducibility of the phenomenon studied (i.e., duct
formation) and the cell phenotype (i.e., response to a given hormone). 
\item \textit{In vitro} 3D models allow researchers to manipulate constraints beyond the range operating \textit{in
vivo}. That is, constraints are determined by the organism and its parts, while in the \textit{in vitro} model the
researcher also plays a direct role in modifying these constraints and parameterizing them. For example, we can
manipulate the rigidity of the mammary gland model to that of bone, and learn how rigidity affects shape beyond the
limits imposed by the organism (Weaver et al. 1995). This type of manipulation revealed that high rigidity inhibits
lumen formation and makes epithelial structures disorganize in a way reminiscent of neoplasms (Paszek et al. 2005).
\item Specific organism level constraints induce the tissue to undergo morphological changes required for proper organ
function at the right time. Hormone action on the mammary gland is an example of this. At the onset of puberty,
estrogen influences the formation of TEBs, the structure at the end of the ducts that invades the stroma and guides
ductal growth until the ductal tree fills the fat pad. Progesterone promotes side-branching and during pregnancy
prolactin facilitates acinar development in preparation for lactation. A classical biochemical interpretation would
reduce the phenomenon to a hormone-receptor interaction triggering signaling pathways inside the cell. However, this
approach does not have the capacity to explain the shape changes resulting from these hormonal influences. From the
tissue perspective, exposure to hormones leads to changes in collagen fiber organization which enable the cells to
generate various epithelial organization patterns. In a hormone-sensitive 3D culture model, epithelial structures
resulting from exposure to estrogen in combination with a progestogen or prolactin were more irregular in shape than
the elongated, smooth structures resulting from exposure to estrogen alone. Consistently, combined hormone treatment
resulted in higher collagen density variability within 20\hspace{0.2em}$\mu $m from the epithelial structure compared
to E2 alone (Speroni et al. 2014) (Figure \ref{fig11}). 
\end{enumerate}

\section{ Conclusions}

We posited that experimental research guided by the global theoretical approach that we are proposing would be different
from that of the prevailing ones which are mostly guided by the metaphors of information, signal and program borrowed
from mathematical information theories (Longo et al. 2012).  Often times, modelers entering biological research treat
biological objects as if they were either physical objects or computer programs. In this issue, we have presented
critiques to these approaches and suggested biological singularities that have to be taken into consideration when
constructing a biological theory (Longo and Soto, this issue, Longo et al. 2015).  On the one hand, the metaphorical
use of information pushes the experimenter to seek causality in terms of discrete structures, namely molecules, in
particular DNA. This view precludes physical “constraints”, like the ones analyzed here, to causally contribute to the
generation and maintenance of the organism unless they are digitally encoded as molecular signs. On the other hand,
biophysical approaches involve hypotheses which are not sound in biology, such as optimization principles in the
behavior of cells. The approach adopted herein is based on two of the principles proposed as foundations for a theory
of organisms, namely, the default state and organizational closure. We explored whether it would be possible and
informative to model epithelial glandular morphogenesis from these biological principles, rather than the usual
procedure of transferring mathematical structures developed for the understanding of physical phenomena into biological
ones or proposing that cells follow a program. It is worth stressing that the former represents true
\textit{mathematical modeling} which is based on the theoretical framework of the discipline to which the modeled
phenomenon pertains, while the latter, properly described as \textit{imitation}, uses principles from one discipline
and applies them to another without a critical appraisal of their theoretical meaning when transported into a different
theoretical context.  

In this initial modeling effort, applying the two principles (default state and constraints leading to closure) were
sufficient to show the formation of ducts and acini. Cells generated forces that were transmitted to neighboring cells
and collagen fibers, which in turn created constraints to movement and proliferation (Figures \ref{fig7} \& \ref{fig8}). Additionally,
constraints to the default state are sufficient to explain ductal and acinar formation, and point to a target of future
research, namely, the inhibitors of cell proliferation and motility which in this mathematical model are generated by
the epithelial cells. Finally, the success of this modeling effort performed as a “proof of principle” opens the
possibility for a step-wise approach whereby additional constraints imposed by the tissue (e.g., additional cell types)
and the organism (e.g., hormones) could be assessed \textit{in silico} and rigorously tested  by \textit{in vitro} and
\textit{in vivo} experiments.

\section*{Acknowledgments}
\begin{sloppypar}
 This work was conducted as part of the research project “Addressing biological organization in the post-genomic era”
which is supported by the International Blaise Pascal Chairs, Region Ile de France (AMS: Pascal Chair 2013). Additional support to AMS was
provided by Grant R01ES08314 from the U. S. National Institute of Environmental Health Sciences and Maël Montévil was
supported by Labex {\textquotedbl}Who am I?{\textquotedbl}, Laboratory of Excellence No. ANR-11-LABX-0071. The funders
had no role in the study design, data collection and analysis, decision to publish, or preparation of the manuscript.
The authors are grateful to Cheryl Schaeberle for her critical input and Clifford Barnes for image acquisition.  The
authors have no competing financial interests to declare.

\end{sloppypar}

\section{References}
\begin{sloppypar}

{\setlength{\parindent}{-20pt}
\setlength{\leftskip}{20pt}
Balinsky, B.I. 1950. On the prenatal growth of the mammary gland rudiment in the mouse. \textit{J. Anat. 84}, 227-235.

Barnes, C., Speroni, L., Quinn, K., Montévil, M., Saetzler, K., Bode-Animashaun, G., McKerr, G., Georgakoudi, I., Downes, S., Sonnenschein, C., Howard, C.V., Soto, A.M. 2014. From single cells to tissues: interactions between the matrix and human breast cells in real time. \textit{PLoS ONE 9}, e93325.

Bizzarri, M., Cucina, A., Palombo, A., Masiello, M.G. 2014. Gravity sensing by cells: mechanisms and theoretical grounds. \textit{Rend. Fis. Acc. Lincei 25 Suppl 1}, S29-S38.

Brisken, C., O'Malley, B. 2010. Hormone action in the mammary gland. Cold Spring Harb. \textit{Perspect. Biol. 2}, a003178.

Dallon, J.C., Evans, E.J., Ehrlich, P. 2014. A mathematical model of collagen lattice contraction. \textit{J. R. Soc. Interface 11}, 20140598.

Dhimolea, E., Maffini, M.V., Soto, A.M., Sonnenschein, C. 2010. The role of collagen reorganization on mammary epithelial morphogenesis in a 3D culture model. \textit{Biomaterials 31}, 3622-3630.

Gould, S.J. 1998. \textit{Leonardo's Mountain of Clams and the Diet of Worms: Essays on Natural History}, 155.

Grant, M.R., Hunt, C.A., Xia, L., Fata, J.E., Bissell, M.J.  2004. Modeling mammary gland morphogenesis as a reaction-diffusion process, in: \textit{Engineering in Medicine and Biology Society, JEMBS '04 26th Annual International Conference of the IEEE}, pp. 679-682.

Guo, C.L., Ouyang, M., Yu, J.Y., Maslov, J., Price, A., Shen, C.Y. 2012. Long-range mechanical force enables self-assembly of epithelial tubular patterns. \textit{Proc. Nat. Acad. Sci. USA 109}, 5576-5582.

Harjanto, D., Zaman, M.H. 2013. Modeling extracellular matrix reorganization in 3D environments. \textit{PLoS One. 8}, e52509.

Iber, D., Menshykau, D. 2013. The control of branching morphogenesis. \textit{Open Biol. 3}, 130088.

Krause, S., Jondeau-Cabaton, A., Dhimolea, E., Soto, A.M., Sonnenschein, C., Maffini, M.V. 2012. Dual regulation of breast tubulogenesis using extracellular matrix composition and stromal cells. \textit{Tissue Eng Part A. 18}, 520-532.

Krause, S., Maffini, M.V., Soto, A.M., Sonnenschein, C. 2008. A novel 3D in vitro culture model to study stromal-epithelial interactions in the mammary gland. \textit{Tissue Eng. Part C Methods 14}, 261-271.

Longo, G., Miquel, P.-A., Sonnenschein, C., Soto, A.M. 2012. Is information a proper observable for biological organization? \textit{Prog. Biophys. Mol. Biol. 109}, 108-14.

Longo, G., Montévil, M. 2011. From physics to biology by extending criticality and symmetry breakings. \textit{Prog. Biophys. Mol. Biol. 106}, 340-347.

------ 2014. \textit{Perspectives on Organisms: Biological Time, Symmetries and Singularities.} Springer, Berlin.

Longo, G., Montévil, M., Sonnenschein, C., Soto, A.M. 2015. In search of principles for a theory of organisms. \textit{J. Biosci. 40}, 955-968.

 Longo, G. \& Soto, A.M. 2016, this issue. Why do we need theories?, \textit{Progress in Biophysics and Molecular Biology}, Available online 4 July 2016, ISSN 0079-6107, \url{http://dx.doi.org/10.1016/j.pbiomolbio.2016.06.005}

Miquel, P.-A.,  Hwang, S.-Y 2016, this issue. From physical to biological individuation, \textit{Progress in Biophysics and Molecular Biology}, Available online 16 July 2016, ISSN 0079-6107, \url{http://dx.doi.org/10.1016/j.pbiomolbio.2016.07.002}

Montévil, M., Mossio, M., Pocheville, A. \& Longo, G. 2016, this issue. Theoretical principles for biology: Variation, \textit{Progress in Biophysics and Molecular Biology}, Available online 13 August 2016, ISSN 0079-6107, \url{http://dx.doi.org/10.1016/j.pbiomolbio.2016.08.005}

Mossio, M.,  Montévil, M.,  Longo, G. 2016, this issue. Theoretical principles for biology: Organization, \textit{Progress in Biophysics and Molecular Biology}, Available online 10 August 2016, ISSN 0079-6107, \url{http://dx.doi.org/10.1016/j.pbiomolbio.2016.07.005}

Neagu, A. 2006. Computational modeling of tissue self-assembly. \textit{Mod. Phys. Lett. 20}, 1217.

Paine, I., Chauviere, A., Landua, J., Sreekuman, A., Cristini, V., Rosen, J., Lewis, MT. 2016. A geometrically-constrained mathematical model of mammary gland ductal elongation reveals novel cellular dynamics within the terminal end bud. \textit{PLoS Comput. Biol. 12}, e1004839.

Paszek, M.J., Zahir, N., Johnson, K.R., Lakins, J.N., Rozenberg, G.I., Gefen, A., Reinhart-King, C.A., Margulies, S.S., Dembo, M., Boettiger, D., Hammer, D.A., Weaver, V.M. 2005. Tensional homeostasis and the malignant phenotype. \textit{Cancer Cell 8}, 241-254.

 Perret, N. \&  Longo, G. 2016, this issue. Reductionist perspectives and the notion of information, \textit{Progress in Biophysics and Molecular Biology}, Available online 22 July 2016, ISSN 0079-6107, \url{http://dx.doi.org/10.1016/j.pbiomolbio.2016.07.003}

Rejniak, K.A., Anderson, A.R. 2008. A computational study of the development of epithelial acini: II. Necessary conditions for structure and lumen stability. \textit{Bull. Math. Biol. 70}, 1450-1479.

Robinson, G.W., Karpf, A.B.C., Kratochwil, K. 1999. Regulation of mammary gland development by tissue interaction. \textit{J. Mammary Gland Biol. Neoplasia 4}, 9-19.

Sonnenschein C. and Soto A.M.  \textit{The Society of Cells: Cancer and Control of Cell Proliferation}. New York: Springer Verlag; 1999.

Sonnenschein, C., Soto, A.M., Michaelson, C.L. 1996. Human serum albumin shares the properties of estrocolyone-I, the inhibitor of the proliferation of estrogen-target cells. \textit{J. Steroid Biochem. Molec. Biol. 59}, 147-154.

 Sonnenschein, C., Soto, A.M., Carcinogenesis explained within the context of a theory of organisms, \textit{Progress in Biophysics and Molecular Biology}, Available online 3 August 2016, ISSN 0079-6107, \url{http://dx.doi.org/10.1016/j.pbiomolbio.2016.07.004}

Soto, A.M.,  Longo, G.,  Montévil, M.,  Sonnenschein, C. 2016, this issue. The biological default state of cell proliferation with variation and motility, a fundamental principle for a theory of organisms, \textit{Progress in Biophysics and Molecular Biology},  Available online 2 July 2016, ISSN 0079-6107, \url{http://dx.doi.org/10.1016/j.pbiomolbio.2016.06.006}

Soto, A.M., Sonnenschein, C. 2005. Emergentism as a default: cancer as a problem of tissue organization. \textit{J. Biosci. 30}, 103-118.

Soto, A.M., Sonnenschein, C. 2011. The tissue organization field theory of cancer: A testable replacement for the somatic mutation theory. \textit{BioEssays 33}, 332-340.

Soto, A.M., Sonnenschein, C., Miquel, P.-A. 2008. On physicalism and Downward Causation in Developmental and Cancer Biology. \textit{Acta Biotheoretica 56}, 257-274.

Speroni, L., Whitt, G.S., Xylas, J., Quinn, K.P., Jondeau-Cabaton, A., Georgakoudi, I., Sonnenschein, C., Soto, A.M. 2014. Hormonal regulation of epithelial organization in a 3D breast tissue culture model. \textit{Tissue Eng. Part C Methods 20}, 42-51.

Tang, J., Enderling, H., Becker-Weimann, S., Pham, C., Polyzos, A., Chen, C.-Y., Costes, S.V. 2011. Phenotypic transition maps of 3D breast acini obtained by imaging-guided agent-based modeling. \textit{Integrative Biology 3}, 408-421.

Tanner, K., Mori, H., Mroue, R., Bruni-Cardoso, A., Bissell, M.J. 2012. Coherent angular motion in the establishment of multicellular architecture of glandular tissues. \textit{Proc. Nat. Acad. Sci. USA 109}, 1973-1978.

van Fraassen, B.C. 1989. \textit{Laws and Symmetry. }Oxford University Press, Oxford.

Vandenberg, L.N., Maffini, M.V., Wadia, P.R., Sonnenschein, C., Rubin, B.S., Soto, A.M. 2007. Exposure to environmentally relevant doses of the xenoestrogen bisphenol-A alters development of the fetal mouse mammary gland. \textit{Endocrinology 148}, 116-127.

Weaver, V.M., Howlett, A.R., Langton-Webster, B., Petersen, O.W., Bissell, M.J. 1995. The development of a functionally relevant cell culture model of progressive human breast cancer. \textit{Seminars in Cancer Biology 6}, 175-184.

}
\end{sloppypar}

\appendix\renewcommand*{\thesection}{\Alph{section}}

\section{Mathematical description of the model}
\label{appendix}

In this appendix, we describe the mathematical model of epithelial morphogenesis in collagen gels that we propose as a
proof of concept for the use of the notion of proliferation with variation and motility as the default state of cells.

A gel is described as a three dimensional material whose properties change over time.

In our simulations, collagen and more generally the system is approximated as a three dimensional lattice with different
dimensions corresponding to the different scalar or vector fields relevant to the system (collagen orientation,
chemical concentrations, presence of cells, etc). The program has a main loop which corresponds to the update of
cellular behaviors. The updates of collagen or chemicals are performed by sub-loops. This implementation is justified
by the assumption that the characteristic speed of processes described by the sub-loops is faster than the cellular
changes. All coefficients are given with respect to the time scale of the main loop. 

Initially, every element of the array has a random collagen orientation with a uniform distribution
and every element is independent of the other. The nutrient layer is set uniformly to $1$ and the inhibitory
layer to $0$. A
single cell is typically put in position $(25,25,25)$ in a $50\times 50 \times 50 $ array. 

\subsection{Collagen and forces}
We only take into account the mean orientations of the fibers. This mean orientation provides a vector
for every element of the lattice, thus defining a vector field $\overrightarrow{C}(x,y,z,t)$. Collagen
orientation vectors have norm $1$. In order to lighten the notation we will keep the normalization of the vectors implicit in this
text. Note that for all intend and purpose, vectors $\overrightarrow{C}(x,y,z,t)$ and $-\overrightarrow{C}(x,y,z,t)$ are equivalent
because we assume that collagen fibers are not oriented; thus, this symmetry is respected in all equations. 

\begin{sloppypar}
Forces are represented by several vector fields 
$\overrightarrow{F}_i(x,y,z,t)$, with $i=x$, $y$, or $z$. $\overrightarrow{F}_x(x,y,z,t)$
corresponds to
the force exerted by the element in position $(x-1,y,z)$ on the element in
position $(x,y,z)$ at time $t$. Note that
according to the principle of reaction of classical mechanics, $-\overrightarrow{F}_x(x,y,z,t)$  is then the force
exerted by the element in position $(x,y,z)$ on the element in position $(x-1,y,z)$. Collagen is
considered as an anisotropic elastic material at short time scales. Collagen orientation is altered at a larger time
scales. We simulate the collagen and force propagation by the finite difference method. 

\end{sloppypar}

 The anisotropy of collagen is modeled by a Young’s modulus which depends on collagen orientation:
 \[\left((1-\alpha)\overrightarrow{C}(x,y,z,t)+\alpha\right)(1-d)\]
 (we use $d=0.1$ and $\alpha=0.5$). The
representation of collagen that we use is a crude macroscopic representation, but we feel that it is a computationally
“lightweight” approach that is sufficient to understand the morphogenesis of the epithelial structures of interest.
Note also that both the mathematical and the biological models that we discuss aim ultimately to understand
morphogenesis \textit{in vivo}. 3D cell cultures are already an experimental
model of this phenomenon. As a result, there is no reason here to aim for a highly detailed representation of collagen.

Cells transmit forces in the same manner than collagen (orientation is then the orientation of the cytoskeleton). Cells
may also exert forces; this is discussed below. However, elements which describe the lumen do not transmit forces.

Collagen changes its orientation when forces are exerted on it, more precisely

\[\overrightarrow{C}(x,y,z,t+1)=\epsilon \overrightarrow{C}(x,y,z,t) + \frac{\Delta t}{F_1\tau_c} \overrightarrow{F}(x,y,z,t)    \]

This equation is valid modulo a normalization and a sign factor. Here, $\epsilon=\pm 1$
 so that the two
vectors have the same orientation (a positive dot product) which account for the fact that collagen orientation is
defined up to a factor $-1$
,as mentioned above. We use a relatively small $\Delta t/F_1\tau_c$  ($0.01$) because
remodeling is a relatively slow process in comparison with mechanical equilibrium. This equation corresponds to an
averaging of the orientation of collagen and of the orientation of the forces at each time step with a weight of the
force that depends on its magnitude.

 In the case of acini formation in Matrigel, we assume that cells cannot exert forces on the extracellular matrix. In
our model, this assumption is equivalent to the assumption that the extracellular matrix cannot transmit forces.

Cells will respond to the stress exerted on it:
\begin{multline*}
 \overrightarrow{F}(x,y,z,t)= \overrightarrow{F}_x(x,y,z,t) + \overrightarrow{F}_y(x,y,z,t) \\+ \overrightarrow{F}_z(x,y,z,t) + \overrightarrow{F}_x(x+1,y,z,t) \\+ \overrightarrow{F}_y(x,y+1,z,t) + \overrightarrow{F}_z(x,y,z+1,t)
\end{multline*}

Note that the different axes of this stress may be compressive or tensile independently. Also the expression above means
that we do not distinguish shear stress from axial stress for the cellular response which is a simplifying
approximation.

We also assume that the magnitude of this vector is a relevant quantity for cells: 

\[F_m(x,y,z,t)=\left\lVert\overrightarrow{F}(x,y,z,t) \right\rVert\]

\subsection{ Chemical layers}
We consider two chemical concentration scalar fields. The first is the “nutrient” layer which is
involved in lumen formation. We model this layer by a diffusion equation with fixed boundary conditions (set to $1$), and we simulate
it by finite differences. Cells consume these nutrients. When the local concentration in nutrient is below a threshold,
the cell dies which leads to lumen formation at the center of structures.

The second chemical component is the inhibitory layer discussed in the text. It is simulated as a chemical that decays
rapidly and diffuses. Its effects are thus confined to the neighborhood of its sources. The sources are cells adjacent
to a lumen. Because of these specific features, extracellular matrix deposition in the same region would have
approximately the same distribution pattern. When this layer is above a threshold in a given element occupied by a
cell, both proliferation and motility are impossible for this cell. Further work will aim to distinguish the properties
of a chemical inhibitor and the properties of the basement membrane components secreted by epithelial cells. 

\subsection{ Cellular proliferation and motility}
\subsubsection{ Cell Movement}
We model cell-cell contact interactions in a macroscopic manner. We consider $n_{nei}(x,y,z,t)$  the number of
neighbors of a cell at position $(x,y,z)$. Note that we normalize the contribution of every neighbor by distances, so that a cell that is a
neighbor with a relative position $(1,1,0)$ counts for  $1/\sqrt{2}$. We also use the distribution of probability $p_{x,y,z,t}$
 with 
\begin{multline*}
   p_{(x,y,z,t)}(i,j,k) \propto \delta(x+i,y+j,z+k,t ) \times \\ \left(n_{nei}\left(x+i,y+j,z+k,t\right) +\frac{0.5}{ \sqrt{i^2+ j^2+k^2}} \right)^2
\end{multline*}

where $(i,j,k) \in \{-1,0,1\}^3 -{(0,0,0)}$ and $\delta(x+i,y+j,z+k,t )$  is $1$ if there are no cell in this position or $0$ otherwise. Note that the counts here ignore the cell
at position $(x,y,z)$
since this cell is the object under influence of the other cells in our uses of this distribution. This distribution
means that the choice of a direction is isotropic for an isolated cell. However, when there are other cells next to it,
a cell will tend to select a direction towards the other cells even though the opposite is still possible.

In the absence of the inhibitor discussed above, cells initiate motion with a probability 

    \[\exp\left(-\frac{F_m(x,y,z,t)}{f_m} -\frac{n_{nei}(x,y,z,t)}{n_{inhib}}\right)\]

Here, $f_m$  is the characteristic magnitude of the inhibition of cell movement by physical forces (we use $f_m=40$). The inhibition
of motility by cell to cell contact is described by $n_{inhib}$ (we use $n_{inhib}=3$). Then, the
possible motion follows a random direction $\overrightarrow{r}$
 with the 
probability distribution $  p_{(x,y,z,t)}$. This motion leads to the new position $(x',y',z')$. The motion is
performed if the following condition is met: 

\[ \frac{n_{nei}(x',y',z',t)+0.5}{n_{nei}(x,y,z,t) +0.5} \prod_{\overrightarrow{v}} \left(\frac{\lvert\overrightarrow{r}.\overrightarrow{v}\rvert+0.25}{\lvert\lvert \overrightarrow{v} \rvert\rvert+0.25}\right)^{1/3}  -X >0\]

where $X$  is a random variable with uniform distribution over $[0,1]$ and $\overrightarrow{v}$ is $\overrightarrow{C}(x,y,z,t)$,$\overrightarrow{C}(x',y',z',t)$ and $\overrightarrow{F}(x',y',z',t)$. We thus assume
that movement is facilitated towards positions where the number of neighboring cells is larger than in the initial
position. Positions which are easier to access correspond to a displacement in the direction of collagen orientation
and local forces. We combine these last effects by considering their geometric average. 

\subsubsection{ Proliferation}
Every cell has an internal variable $\theta$ which corresponds
to the progress of its cell cycle. Cells may only proliferate when $\theta$ reaches $1$. In the lack of
the inhibitor mentioned above, $\theta$ is incremented by $\Delta \theta=\Delta t / \tau_p$ (we use $\Delta \theta=1/6$) where $\Delta t$ is the duration of
an iteration of the program and $\tau_p$is the time required for proliferation when there are no constraints (maximal
proliferation). 

When $\theta$ reaches $1$, the cell attempts to proliferate with a probability $\exp(-F_m(x,y,z,t)/f_p)$ where $f_p$ is a
characteristic magnitude of the inhibition of proliferation by physical forces (we use $f_p=80$). The direction
in which the cell tries to proliferate is given by a combination of $F(x,y,z,t)$ and a random
vector $\overrightarrow{r}$ generated
on the basis of the probability distribution $p_{x,y,z,t}$ described for
motility. This random vector leads to the position $(x',y',z')$  that is a neighbor
of the initial position. More precisely the direction of the proliferation attempt is given by:

\[ \epsilon_1 \overrightarrow{F}(x,y,z,t)+\epsilon_2 w_c \overrightarrow{C}(x,y,z,t) +w_r n_{nei}(x',y',z',t) \overrightarrow{r} \] 

Let us recall that $C$ in an element occupied by a cell represents the orientation of the cytoskeleton. $w_c$ is the impact of
the cytoskeleton orientation on the direction of proliferation; we use $w_c=1$. $w_r$  is the weight of
the random factor. Of course, all these weights are relative to the magnitude of the force $F$.  $\epsilon_i$  are random
coefficients with values which are either  $1$ or $-1$. If the resulting
direction in which the new cell would appear is occupied, the process is aborted and a proliferation attempt is
performed at the next time step. Cells never stop trying to proliferate, but they will stop proliferating when strongly
constrained. 

\subsection{Cellular forces}
Every cell has an internal variable $\overrightarrow{\phi}(t)$  (a normalized
vector) which corresponds to the retention of the direction of the former forces it exerted. This vector corresponds to
the orientation in which the cell “chooses” to preferentially pull or push. The new orientation is: 

\begin{multline*}
 \overrightarrow{\phi}(t+1) = 10 \overrightarrow{\phi}(t) +2X+ \overrightarrow{F}(x,y,z,t)\\
 -0.5 F_m(x,y,z,t)^2 \overrightarrow{F}(x,y,z,t) +\sum_ {\overrightarrow{v} \in A}  \overrightarrow{v}.\overrightarrow{C}(x,y,z,t) \overrightarrow{v} 
\end{multline*}

where the sum is over the set $A$ of directions to
neighbor positions which are occupied by cells. $X$ is a random vector
of norm $1$. Note
that cells react to forces by amplifying small forces and opposing strong forces which we model by the polynomial
response above.

The force that a cell at position $(x,y,z)$ can exert on an
adjacent element at $(x,y,z)+\overrightarrow{v}=(x',y',z')$
 is: 

\[  \left(\left\lvert\overrightarrow{v}.\overrightarrow{C}(x,y,z,t)\right\rvert + \left\lvert\overrightarrow{C}(x,y,z,t).\overrightarrow{C}(x',y',z',t)\right\rvert\right) F \overrightarrow{v}  \]

where  $F$
 is either  $F_{cell}$ 
 if the adjacent position $(x',y',z')$ is a cell or $F_{col}$
  if this position
is occupied by collagen. We use, for example,  $F_{cell}=4$ and $F_{col}=3$.

Then the force that the cell attempts to exert is finally:

\[  \left(\left\lvert\overrightarrow{v}.\overrightarrow{C}(x,y,z,t)\right\rvert + \left\lvert\overrightarrow{C}(x,y,z,t).\overrightarrow{C}(x',y',z',t)\right\rvert\right) F \overrightarrow{v}.\overrightarrow{\phi}(t)  \overrightarrow{v}\]

The forces exerted by cells are added to the forces present in the collagen gel, and the new mechanical system is then
simulated. Note that the force effectively exerted may be different because the mechanical simulation of the system
leads to the elimination of systemic inconsistencies. For example in our model it is not possible for the cell to exert
forces on the lumen.

\section{Supplementary videos}

The videos will open by clicking on the images, alternatively they may be found at:
\begin{sloppypar}
 \href{http://montevil.theobio.org/en/content/videos-modeling-mammary-organogenesis-biological-first-principles-cells-and-their-physical}{http://montevil.theobio.org/en/content/videos-modeling-mammary-organogenesis-biological-first-principles-cells-and-their-physical}

\end{sloppypar}

\hypersetup{pdfborder={0 0 1}, urlcolor=.}

\makeatletter
 \def\Hy@colorlink#1{%
      \begingroup
      \HyColor@UseColor#1%
    }    \def\Hy@endcolorlink{\endgroup}%

\makeatother

\begin{myfloat}[h]
\centering
 \href{http://montevil.theobio.org/sites/montevil.theobio.org/files/videos_morphogenesis/video1.avi}{\includegraphics[scale=0.25]{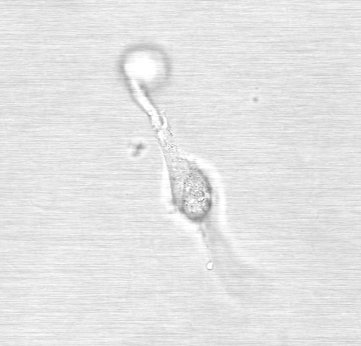}} 
\caption{Projections of breast epithelial cells seeded in a fibrilar matrix.  Cells emit projections in all directions
soon after seeding. These cell projections are involved in collagen organization. 
}\label{vid1}
\end{myfloat}

\begin{myfloat}
\centering
 \href{http://montevil.theobio.org/sites/montevil.theobio.org/files/videos_morphogenesis/video2.avi}{\includegraphics[scale=0.25]{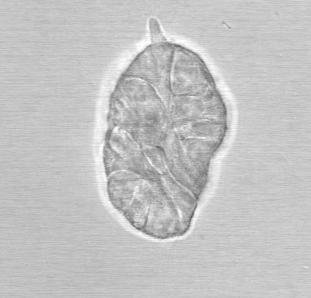}}
\caption{Breast epithelial cells forming an acinus in a non-fibrilar matrix at day 4. Cells display limited motility and
emit only short projections into the matrix. Cells rotate and divide resulting in the formation of an acinus, a sphere
with a central lumen. 
}\label{vid2}
\end{myfloat}

\begin{myfloat}
\centering
 \href{http://montevil.theobio.org/sites/montevil.theobio.org/files/videos_morphogenesis/video3.avi}{\includegraphics[scale=0.15]{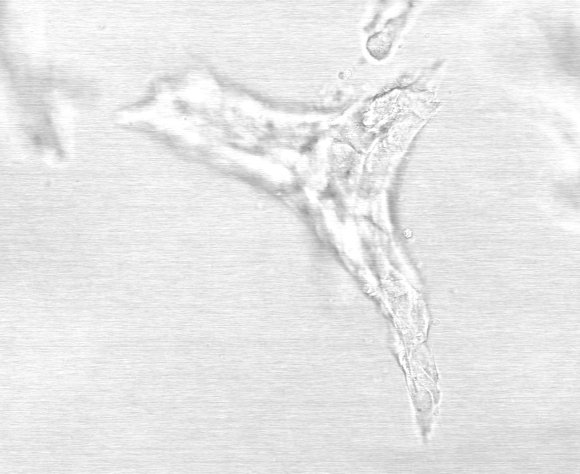}}
\caption{ Branching duct at day 7 of culture. A cell detaches from the main structure and is incorporated back into the
structure. 
}\label{vid3}
\end{myfloat}

\begin{myfloat}
\centering
 \href{http://montevil.theobio.org/sites/montevil.theobio.org/files/videos_morphogenesis/video4.avi}{\includegraphics[scale=0.15]{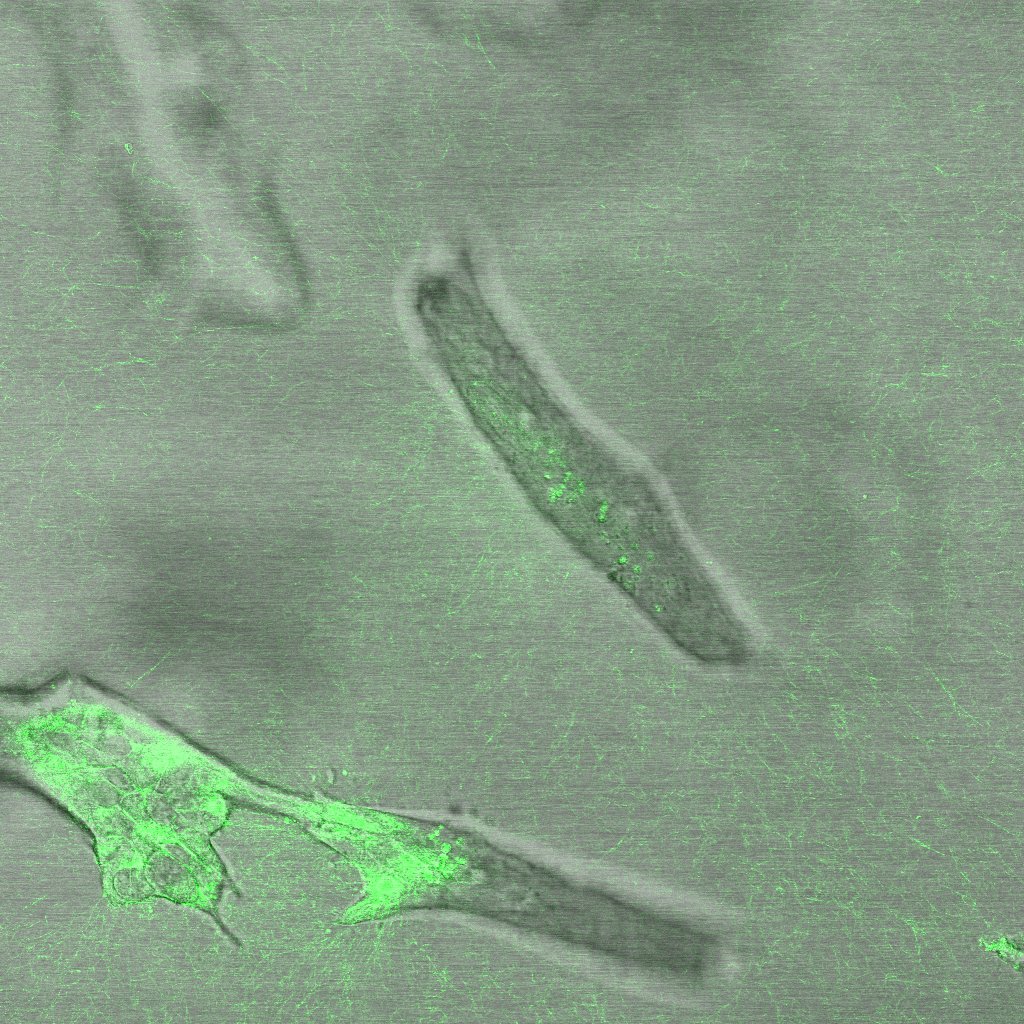}}
\caption{Collagen fibers and breast epithelial structures after 6 days in culture. Cells organize collagen in a collagen
only matrix and the collagen bundles (green) facilitate the merging of epithelial structures.
}\label{vid4}
\end{myfloat}

\begin{myfloat}
\centering
 \href{http://montevil.theobio.org/sites/montevil.theobio.org/files/videos_morphogenesis/video5.avi}{\includegraphics[scale=0.25]{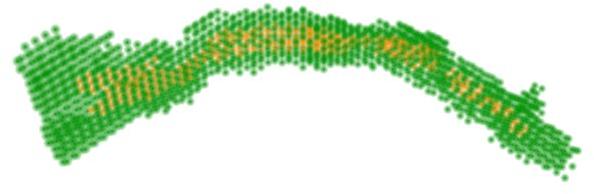}}
\caption{Formation of a duct by the mathematical model. Cells are represented in green and the ductal lumen in
orange in 3 dimensions.  Cells organize collagen over time by exerting forces; in turn, collagen constrains cell
proliferation and motility. These interactions lead to the emergence of a main direction of growth.
}\label{vid5}
\end{myfloat}

\begin{myfloat}
\centering
 \href{http://montevil.theobio.org/sites/montevil.theobio.org/files/videos_morphogenesis/video6.avi}{\includegraphics[scale=0.25]{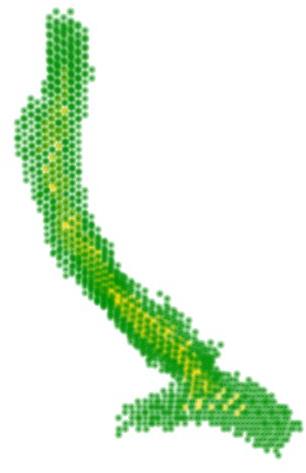}}
\caption{Formation of a duct by the mathematical model. Cells are represented in green and the ductal lumen in
orange in 3 dimensions.  Cells organize collagen over time by exerting forces; in turn, collagen constrains cell
proliferation and motility. These interactions lead to the emergence of a main direction of growth.
}\label{vid6}
\end{myfloat}

\begin{myfloat}
\centering
 \href{http://montevil.theobio.org/sites/montevil.theobio.org/files/videos_morphogenesis/video7.avi}{\includegraphics[scale=0.25]{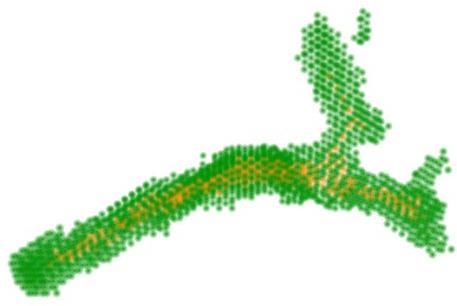}}
\caption{Formation of branching ducts by the mathematical model. Cells are represented in green and the ductal
lumen in orange in 3 dimensions. Ducts branch spontaneously in our model (see also the end of video 5 and 6). For these
simulations, we reduced the range of the inhibitor which increases the odds of branching.
}\label{vid7}
\end{myfloat}

\begin{myfloat}
\centering
 \href{http://montevil.theobio.org/sites/montevil.theobio.org/files/videos_morphogenesis/video8.avi}{\includegraphics[scale=0.25]{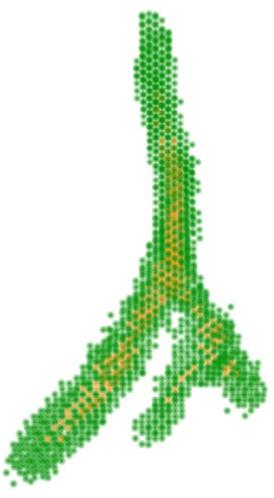}}
\caption{Formation of branching ducts by the mathematical model. Cells are represented in green and the ductal
lumen in orange in 3 dimensions. Ducts branch spontaneously in our model (see also the end of video 5 and 6). For these
simulations, we reduced the range of the inhibitor which increases the odds of branching.
}\label{vid8}
\end{myfloat}

\begin{myfloat}
\centering
 \href{http://montevil.theobio.org/sites/montevil.theobio.org/files/videos_morphogenesis/video9.avi}{\includegraphics[scale=0.25]{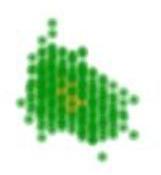}}
\caption{Formation of an acinus by the mathematical model. Cells are represented in green and the lumen in orange in 3
dimensions. Here, the interactions between the ECM and the cells were removed in order to mimic the effect of Matrigel.
 In this condition, cells proliferate and move in an isotropic manner leading to the formation of a rounded structure. 
}\label{vid9}
\end{myfloat}
\end{document}